\documentclass[%
 reprint,
superscriptaddress,
 amsmath,amssymb,
 aps,
]{revtex4-2}

\usepackage{graphicx}
\usepackage{dcolumn}
\usepackage{bm}

\usepackage{xcolor}

\begin{document}
\preprint{APS/123-QED}
\title{Plasmon lifetime enhancement in a bright-dark mode coupled system}

\author{Bilge Can Yildiz}
\email{bilge.yildizkarakul@tuni.fi}
\affiliation{Department of Physics, Middle East Technical University, 06800 Ankara, Turkey}
\affiliation{Institute of Nuclear Sciences, Hacettepe University, 06800 Ankara, Turkey}
\affiliation{Faculty of Engineering and Natural Sciences, Photonics Laboratory, Tampere University, 33720 Tampere, Finland}

\author{Alpan Bek}
\affiliation{Department of Physics, Middle East Technical University, 06800 Ankara, Turkey}

\author{Mehmet Emre Tasgin}
\affiliation{Institute of Nuclear Sciences, Hacettepe University, 06800 Ankara, Turkey}

\date{\today}

\begin{abstract}
Metallic nanoparticles can localize the incident light to hotspots as plasmon oscillations, where the intensity can be enhanced by up to four orders of magnitude. Even though the lifetime of plasmons are typically short, it can be increased via interactions with quantum emitters, e.g. spaser nano-lasers. However, molecules can bleach in days. Here, we study the lifetime enhancement of plasmon excitations due to the coupling with longer lifetime dark plasmon modes. We apply an analytical model based on harmonic oscillators to demonstrate that a coupled system of bright and dark plasmon modes decays more slowly compared to the bright mode alone. Furthermore, exact solutions of the 3D Maxwell equations, i.e. FDTD, demonstrate that the lifetime of the coupled system significantly increases at the hotspot, which is not predictable by far-field response. The decay of the overall energy of such a coupled system, which can be extracted from experimental absorption measurements, is substantially different, compared to the decay of the hotspot field. This observation enlightens the plasmonic applications, where the hotspot intensity enables the detection of the optical responses.
\end{abstract}

\maketitle

\section{\label{sec:level1}Introduction}

Generation of strong local electromagnetic fields at the nanoscale is one of the major objectives in plasmonics. Resonant interaction of metal nanostructures (MNSs) with incident optical light provides localization of electromagnetic fields with up to ten thousand times higher intensity as compared to the incident field \cite{Stockman2011Nanoplasmonics:Future}. Achieving large electromagnetic field enhancements opens up development of fundamentally new metal-based subwavelength optical elements with broad technological potential in biological sensing \cite{Anker2008BiosensingNanosensors, Liu2015SurfacePlatforms, Mejia-Salazar2018PlasmonicBiosensing}, optical nanoantennas \cite{Dregely20113DArray, Akselrod2014ProbingNanoantennas, Savaliya2017TunableMaterials, Yildiz2019}, subwavelength optical imaging \cite{Fang2005Sub-Diffraction-LimitedSuperlens, Maier2007l, Kawata2009}, fluorescence enhancement \cite{Pompa2006, Bharadwaj2007SpectralEnhancement, Liu2013, Carreno2016ResonanceNanoparticle, Hsu2017Plasmon-CoupledTransfer}, plasmonic metamaterials \cite{Garcia-Vidal2005, Zheludev2011, Naik2013}, and nonlinear plasmonics \cite{Kawata2013, Zhang2015, Yildiz2015EnhancedNanostructures, Singh2016EnhancementPaths, Drachev2018EngineeredArray, Postaci2018SilentIntensities}. Despite remarkably enhanced amplitude, plasmon lifetime is typically short, most importantly due to radiative damping for MNSs larger than 20 nm \cite{Melikyan2004OnNanoparticles}. Lifetime of localized surface plasmons have been of interest in recent years \cite{Kirakosyan2016SurfaceNanoshells, Mahan2018LifetimePlasmons, DiVece2018VeryParticles, Chapkin2018LifetimeLimit.}.  

Noginov et. al. 2009, demonstrate a narrowing in the emission spectrum of a spacer, composed of a gold nanoparticle core placed in a dye-doped silica shell \cite{Noginov2009DemonstrationNanolaser}. Coupling between the gold nanoparticle and the dye molecules enables the system to response with a decay rate of the molecular excited level, as a result of resonant energy transfer from excited molecules to surface plasmons. Lifetime extension due to coupling between plasmonic oscillators and quantum emitters is also theoretically demonstrated \cite{Tasgin2013MetalLifetime}, where the path interference effects are utilized. In plasmonic solar cell applications, improved lifetime of plasmons play a very important role on the device efficiency, as light is carried along the photovoltaic structure for longer durations. On the other hand, obtaining lifetime enhancement without using quantum structures would simplify nano-fabrication. Quantum emitters are not as convenient as larger particles in the range from a few to a few hundred nanometers to be used in optical systems \cite{Stockman2010Dark-hotResonances}. Besides, molecules have limited exposure times. Instead of an atomic state, a dark plasmonic state could be advantageous in the sense that interaction between a plasmonic mode and an atom cannot be as strong as interaction between two plasmonic modes, due to their overlapping of the spatial extensions.  

\begin{figure}
\includegraphics[width=\linewidth]{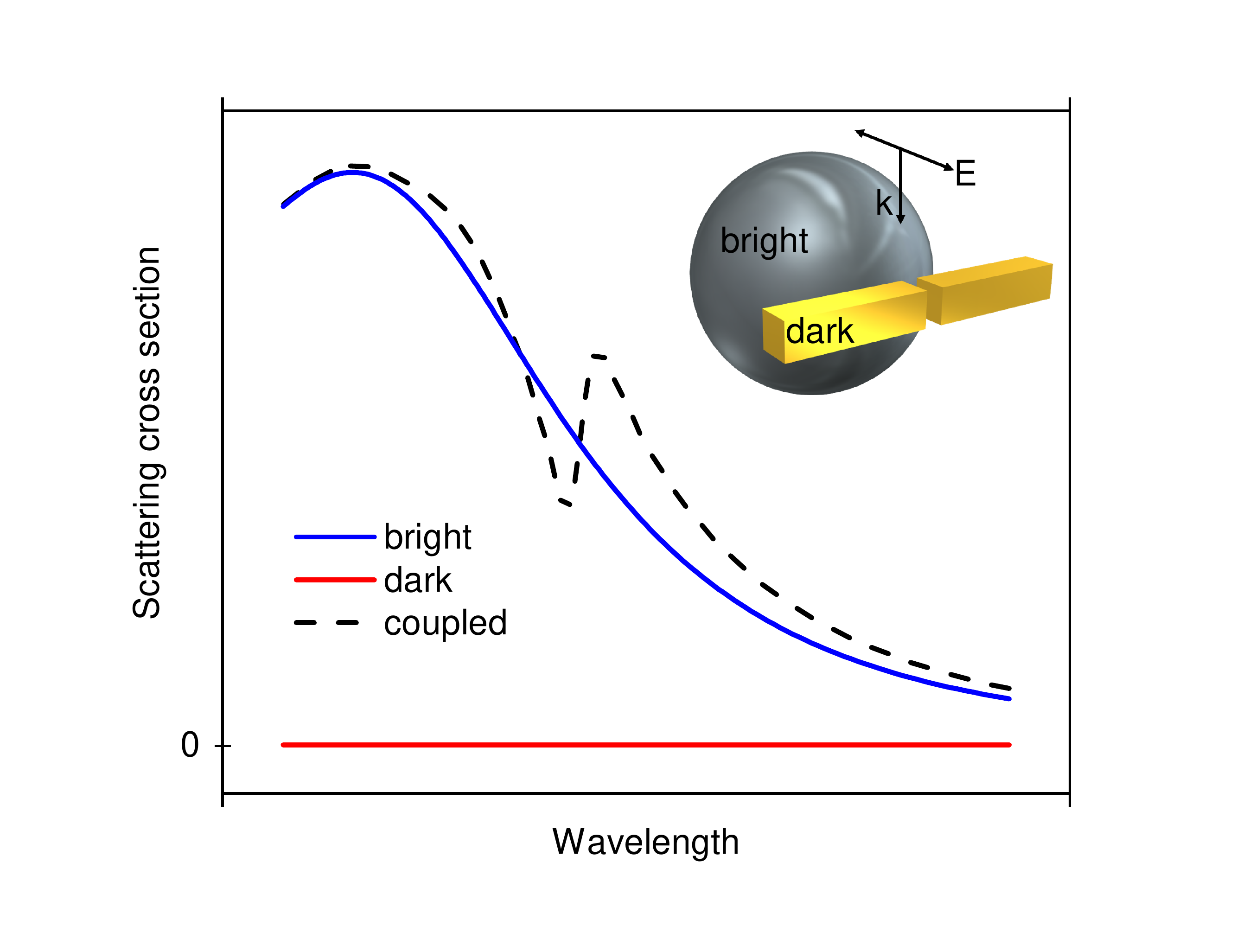}
\caption{\label{fig:fig1} Description of the coupled plasmonic system: A metal nanosphere supporting a bright plasmon mode and a metal nanorod dimer supporting a dark plasmon mode, which cannot be excited directly by the incident light. The graph shows representative optical responses of the individual and coupled systems; the dark mode appears when it is coupled to the bright mode.}
\end{figure}

We investigate the effects of coupling between dark and bright plasmon modes to the plasmon lifetime. We consider a system of two MNSs, one of which supports a plasmonic dark state. A plasmonic dark state is a long-lived excitation that cannot be excited by linearly polarized field, but can couple to (be excited by) near-field polarization of an excited bright plasmon mode \cite{Gomez2013ThePlasmonics, Panaro2014DarkApproach, Gao:18}. When a system consists of plasmonic structures with modes resonant to a source field, this field can be confined into nanoscale dimensions, and interact with long-lived dark states. We consider such a system, illustrated in Fig. \ref{fig:fig1}, where the two plasmonic oscillators are coupled. Although we consider here a scheme where the dark mode belongs to a second MNS to simplify the FDTD simulations, the dark mode could also belong to the same MNS.

In this paper, we apply an analytical theoretical model based on harmonic oscillators, describing the oscillation dynamics of the two coupled plasmon modes. We demonstrate that the lifetime of the coupled system is enhanced compared to the uncoupled one, and the enhancement is the highest at an optimum phenomenological coupling constant, $f$, which is related with the distance between the two plasmonic structures. We numerically monitor the decay of the near-field of the coupled system for a range of inter-structure distance by finite difference time domain (FDTD) \cite{Taflove2005ComputationalMethod} method. The lifetime of the system for changing values of the inter-structure distance exhibits a similar behavior to the one obtained by the theoretical model. Furthermore, FDTD simulations show that the decay time of the electric field at the gap between the MNSs extends significantly, compared to the decay time of the short-lifetime MNS's near-field in the absence of the long-lifetime MNS.  

\section{Analytical Model}

The model system consists of two interacting harmonic oscillators, corresponding to two plasmon modes, supported by two interacting MNSs, in the weak coupling regime. One of the MNSs supports a bright plasmon mode that is resonantly excited by the incident harmonic field. The lifetime of this driven plasmon mode is typically short. The second MNS supports a dark plasmon mode with a vanishing dipole moment, which may arise from the anti-bonding interaction of dipolar resonances on that MNS \cite{C6NR03806A}. Dark plasmon modes cannot be excited by the incident harmonic fields due to weak coupling to the far-field and are typically long-lived. When the two MNSs are brought together, the near-field of the driven MNS excites the dark plasmon mode of the second MNS and the overall system displays hybridized plasmon resonances with modified lifetimes. The individual bright and the dark plasmon modes (in the absence of coupling) are described by amplitudes $\alpha_i^{(0)}$, and associated with natural frequencies $\omega_i$, and damping rates, $\gamma_i$, where $i=1,2$, respectively. The uncoupled eigenmodes are

\begin{equation}
    \alpha_i^{(0)}(t) \propto e^{-(i\omega_i+\gamma_i)t}.
\label{eq:eq1}
\end{equation}

\noindent
The total number of plasmons of the system, in the absence of coupling is determined by $\alpha_1^{(0)}(t)$, since the other mode is not excited without coupling. When the two particles are brought together, the modes are hybridized and there are contributions from the oscillations on both of the particles. The interaction between the oscillators is defined by a parameter $f$, in dimension of frequency. Physically, $f$ quantifies the strength of the coupling between the near-fields of the two plasmon modes. It is proportional to an overlap integral which runs over the spatial overlap of the bright and dark modes \cite{Tasgin2018FanoResponse}. A non-zero $f$ corresponds to a case when the two MNSs are brought together; close enough for the polarization fields induced on the structures to overlap and hence interact. Therefore $f$ can be related with the physical distance between the two MNSs, and it approaches to zero as they diverge away from each other.

The interaction between particles is expected to introduce path interferences at certain frequencies. When there is a nonzero coupling between the structures, the solutions of the oscillations are hybridized \cite{Nordlander2004PlasmonDimers}; being in the form of linear combinations of the two new oscillation modes, as,

\begin{equation}
    \alpha_i(t) \propto e^{h_{i,1}}+e^{h_{i,2}}, 
\label{eq:eq2}
\end{equation}

\noindent where functions appearing on the exponents are functions of the oscillation parameters, i.e., $h=h(\omega_1,\omega_2,\gamma_1,\gamma_2,f,t)$.  The total number of plasmons in the coupled system is defined as,

\begin{equation}
    N(t)=|\alpha_1(t)|^2+|\alpha_2(t)|^2.
\label{eq:eq3}
\end{equation}

\noindent
The lifetime of the induced plasmon oscillations, $\tau$, can be defined as the mean value of the time weighted over plasmon intensity, as follows,

\begin{equation}
    \tau = \frac{\int_0^{\infty }tN(t)dt}{\int_0^{\infty }N(t)dt}.
\label{eq:eq4}
\end{equation}

\noindent
To find an expression for the Eq. (\ref{eq:eq4}), we start with writing the Hamiltonian of the coupled system in the Heisenberg picture as follows,

\begin{equation}
\begin{split}
\hat{H} & = \hbar \omega_1 \hat{a}_1^{\dagger} \hat{a}_1 + \hbar \omega_2 \hat{a}_2^{\dagger} \hat{a}_2 \\
& + \hbar (f \hat{a}_1^{\dagger} \hat{a}_2 + f^* \hat{a}_2^{\dagger} \hat{a}_1)+ i \hbar (\hat{a}_1^{\dagger} \varepsilon_p e^{-i\omega t}),
\label{eq:eq5}
\end{split}
\end{equation}

\noindent
where $\hat{a}_1$ ($\hat{a}_1^{\dagger}$) and $\hat{a}_2$ ($\hat{a}_2^{\dagger}$) are the annihilation (creation) operators for the collective plasmon excitations in the driven and the attached oscillators, corresponding to bright and dark modes. Since properties like entanglement \cite{Sun2008LambModes, Yannopapas2009Plasmon-InducedNanostructures, Qurban2018EntanglementNanoring} are not of interest, they will soon represent the amplitudes of the associated plasmon oscillations $(\hat{a}_i \to \alpha_i)$ and the problem will be reduced to a classical problem. $f$ is the coupling matrix element between the polarization field induced by $\hat{a}_1$ and $\hat{a}_2$. The first and the second terms on the right hand side of Eq. (\ref{eq:eq5}) are the energy operators of the two oscillation modes. The third term corresponds to the interaction energy, and the last term corresponds to the coupling of the pump to the driven oscillator, $\hat{a}_1$. 

\begin{figure}
\includegraphics[width=\linewidth]{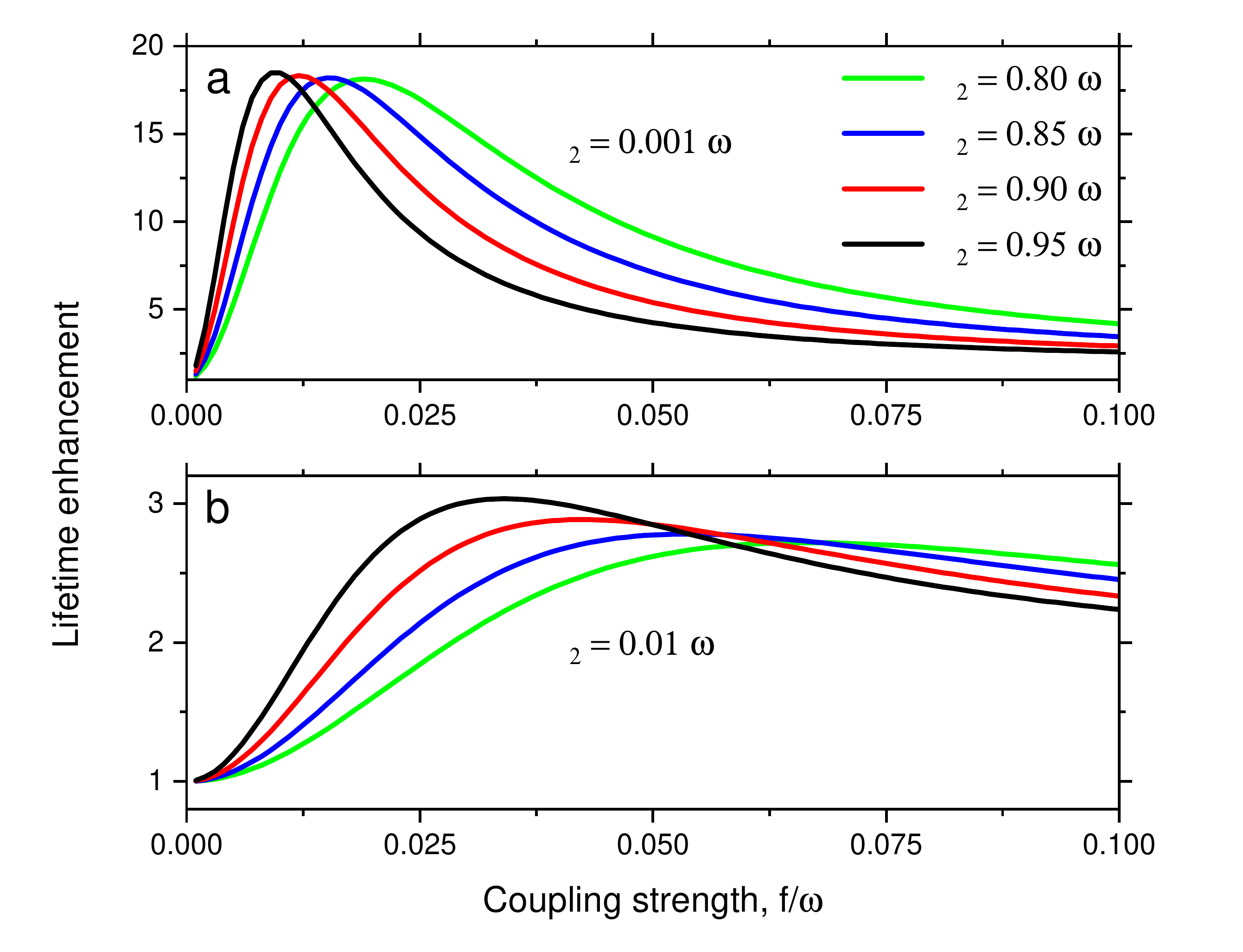}
\caption{\label{fig:fig2}Normalized lifetime of the coupled system, defined by Eq. (\ref{eq:eq4}), as a function of the coupling constant, $f$, for (a) $\gamma_2=0.001 \omega$, and (b) $\gamma_2=0.01 \omega$. In both cases; $\omega_1=1.0 \omega$, and $\gamma_1=0.1 \omega$.}
\end{figure}

Equations of motion for the coupled plasmon modes are obtained using the Heisenberg equation and the damping rates are plugged in. The driving force term is removed, since the decay time of the coupled system such that the driven oscillator is initiated from the excited state, is of the interest. The operators are replaced with their eigenvalues. The equations of motion describing the dynamics of the coupled plasmonic oscillator system are the following,
\begin{eqnarray}
\dot{\alpha}_1 = -(i \omega_1+\gamma_1) \alpha_1 - if \alpha_2, \label{eq:eq6}\\
\dot{\alpha}_2 = - if^* \alpha_1 - (i \omega_2+\gamma_2) \alpha_2. \label{eq:eq7}
\end{eqnarray}

\noindent
Equations (\ref{eq:eq6}) and (\ref{eq:eq7}) form the eigenvalue equation of the system, with the following eigenvalues, $\lambda_{1,2}$, and eigenvectors, $\alpha^{(1),(2)}$,

\begin{eqnarray}
\lambda_{1,2} = \frac{1}{2} [-a-b \pm \sqrt{(a-b)^2-4f^2}], \label{eq:eq8} \\
\alpha^{(1),(2)}  = \begin{pmatrix}
\frac{-if}{\sqrt{f^2+(a+\lambda_{1,2})^2}} \\
\frac{a+\lambda_{1,2}}{\sqrt{f^2+(a+\lambda_{1,2})^2}}
 \end{pmatrix}, \label{eq:eq9}
\end{eqnarray}

\noindent
where,

\begin{equation}
\begin{split}
a & = (i \omega_1+\gamma_1), \\
b & = (i \omega_2+\gamma_2),
\label{eq:eq10}
\end{split}
\end{equation}

\noindent
and $f$ is assigned to be real to reduce the degree of freedom with the purpose of obtaining the simplest and the quickest evidence of lifetime enhancement. The general solution is found as,

\begin{equation}
    \begin{split}
        \alpha_1(t) = C_1 \frac{-if}{\sqrt{f^2+(a+\lambda_1)^2}} e^{\lambda_1t} \\ 
     + C_2 \frac{-if}{\sqrt{f^2+(a+\lambda_2)^2}} e^{\lambda_2t}, \label{eq:eq11}\\
    \end{split}
\end{equation}

\begin{equation}
    \begin{split}
         \alpha_2(t) = C_1 \frac{a+\lambda_1}{\sqrt{f^2+(a+\lambda_1)^2}} e^{\lambda_1t} \\ 
     + C_2 \frac{a+\lambda_2}{\sqrt{f^2+(a+\lambda_2)^2}} e^{\lambda_2t}, \label{eq:eq12}
    \end{split}
\end{equation}

\noindent
where $C_1$ and $C_2$ are the coefficients to be determined according to the initial conditions. For this problem, the incident field is introduced and drives the plasmonic mode that is supported by the MNS with shorter lifetime, and then it is turned off. It is assumed that, initially there are only $\alpha_1$ oscillations. So the initial conditions are given by,

\begin{equation}
    \alpha_1(0)=1, \; \;  \alpha_1(0)=0.
    \label{eq:eq13}
\end{equation}

\noindent
Applying the initial conditions given in Eq. (\ref{eq:eq13}), the solutions are found as,

\begin{equation}
    \alpha_1(t)=\frac{a+\lambda_2}{\lambda_2-\lambda_1}e^{\lambda_1t} - \frac{a+\lambda_2}{\lambda_2-\lambda_1}e^{\lambda_2t},
    \label{eq:eq14}
\end{equation}

\begin{equation}
    \begin{split}
        \alpha_2(t)= - \frac{(a+\lambda_1)(a+\lambda_2)}{if(\lambda_2-\lambda_1)}e^{\lambda_1t} \\
        + \frac{(a+\lambda_1)(a+\lambda_2)}{if(\lambda_2-\lambda_1)}e^{\lambda_2t}.
        \label{eq:eq15}
    \end{split}
\end{equation}

\noindent
Substituting these into Eq. (\ref{eq:eq3}), the lifetime of the system of the two coupled oscillators is obtained from the analytical solutions of the integrations over time in Eq. (\ref{eq:eq4}). Figure \ref{fig:fig2} shows the lifetime enhancement with respect to the coupling strength, $f$, for several different $\omega_{1,2}$ and $\gamma_{1,2}$ sets. The lifetime enhancement is defined by the lifetime of the coupled system, that is calculated by Eq. (\ref{eq:eq4}), divided by the lifetime of the uncoupled system, where the MNS, supporting the dark plasmon mode is not present. In both calculations, the resonance frequency of the bright mode is $\omega_1=1.0\omega$, and the damping rate $\gamma_1=0.1\omega$. Lines with different colors show the cases where $\omega_2=0.80\omega,0.85\omega,0.90\omega$ and $0.95\omega$. In Fig. \ref{fig:fig2}a, the damping rate of the dark mode is $\gamma_2=0.001\omega$, and in Fig. \ref{fig:fig2}b, it is $\gamma_2=0.01\omega$. These numbers make much sense when considering a particular excitation wavelength and scaling all other parameters accordingly. One can consider the excitation frequency to correspond to, for example, 500 nm. Then, the resonance wavelength of the driven mode is at 500 nm, too. The resonance wavelength of the coupled mode is set to be in between 525 nm - 625 nm. The damping rate of the driven mode corresponds to a lifetime of 1.3 fs, whereas the damping rates of the long-lifetime plasmon mode correspond to 130 fs (Fig. \ref{fig:fig2}a) and 13 fs (Fig. \ref{fig:fig2}b), respectively. The selected lifetime of the bright mode (1.3 fs) is typical for a lossy plasmonic mode. Nevertheless, typical plasmon lifetime can vary more than a factor of 10 for different nanoparticle materials and geometries \cite{Andreas2016}. Considering that dark plasmon modes do not possess radiation damping, the decay of these modes are determined mainly by Ohmic losses, resulting in much sharper resonances than that of bright modes. \cite{doi:10.1021/nn102166t, Verellen:11}. 

We observe that when the ratio of the damping rates of the two plasmon modes is 100 (Fig. \ref{fig:fig2}a), which is a rather extreme case, the maximum lifetime enhancement is achieved at $f=0.01-0.02$ for different detunings $(\omega_1-\omega_2)$. When the ratio of the damping rates is 10 (Fig. \ref{fig:fig2}b), which could be typical for dark and bright plasmon modes supported by metal nanostructures, the maximum lifetime enhancement is achieved at $f=0.03-0.06$, compared to the case shown in Fig. \ref{fig:fig2}a. As the detuning of the resonance frequencies gets larger, stronger coupling is required to achieve the same lifetime enhancement, in both cases. The maximum lifetime enhancement is more than an order of magnitude in Fig. \ref{fig:fig2}a, and it is around 3-fold in Fig. \ref{fig:fig2}b. In both cases, $f$ gets an optimum value. That means the lifetime of the coupled system is maximum for some optimum separation, $d$, between the MNSs. For the values of $f$ larger than its optimum,  the lifetime enhancement saturates. This corresponds to a case where the MNSs are getting closer. Since the theoretical model examines the interactions in the weak coupling regime, we do not consider the physical situations, where the MNSs are very close to each other, i.e., strongly coupled. For all the values of $f$ ($d$) smaller (larger) than its optimum, the lifetime enhancement exceeds unity, and it is unity when $f=0$, that is when the MNSs are no longer coupled ($d \rightarrow \infty$).

\begin{figure}
\includegraphics[width=\linewidth]{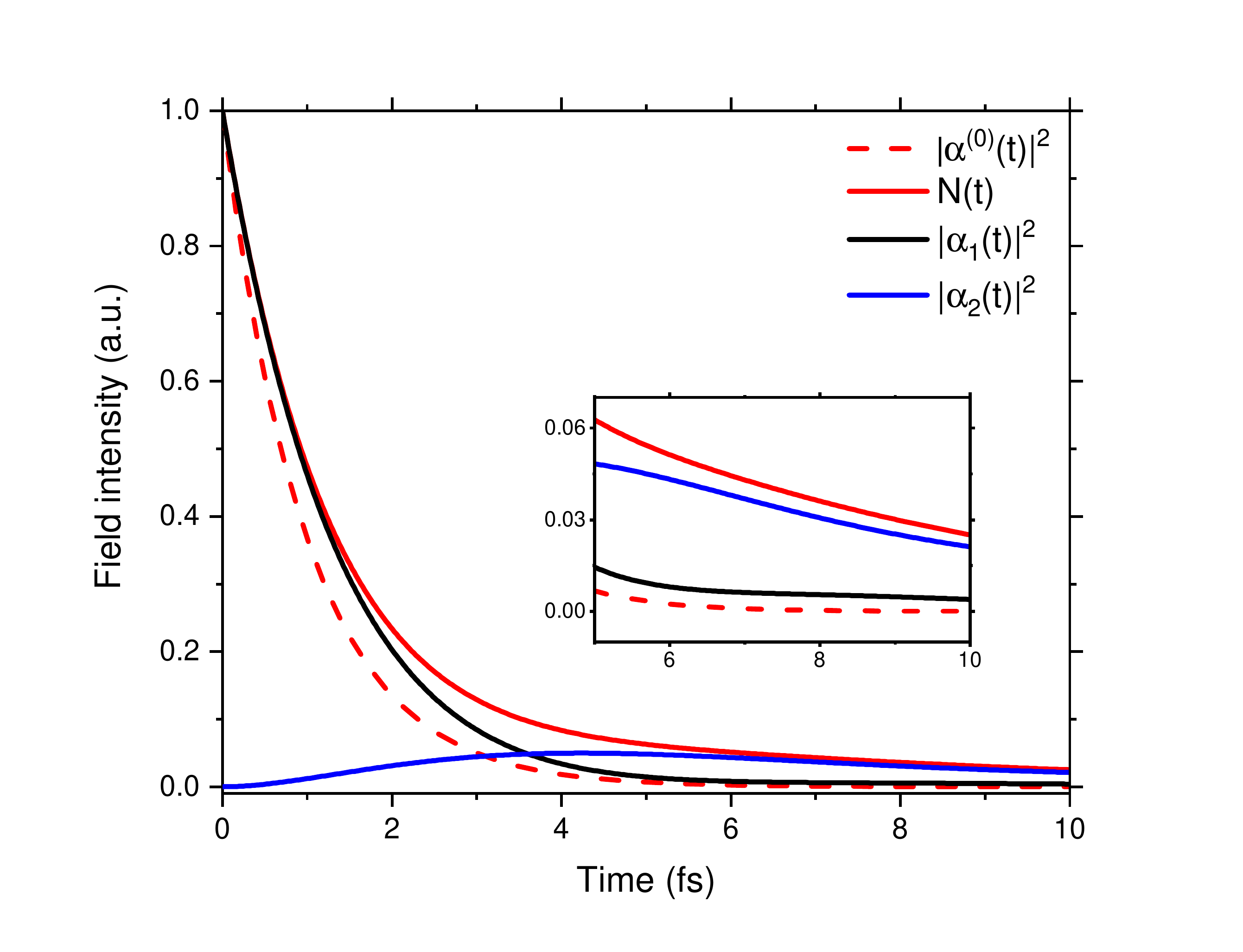}
\caption{\label{fig:fig3}Plasmon intensity for the uncoupled (dashed line) and coupled plasmon modes for $\omega_1=1.0\omega$, $\omega_2=0.9\omega$, $\gamma_1=0.1\omega$, $\gamma_2=0.01\omega$, and $f=0.025\omega$, where the time scale is determined by considering an excitation frequency $\omega$, corresponding to a wavelength of 500 nm. In the inset, the axes are zoomed in at a later time scale.}
\end{figure}

Figure \ref{fig:fig3} shows how the plasmon intensities decay in time, for coupled (solid lines) and uncoupled (dashed line) cases. These field decays are obtained for $\omega_1=1.0\omega$, $\omega_2=0.9\omega$, $\gamma_1=0.1\omega$, $\gamma_2=0.01\omega$, and $f=0.025\omega$, where the time scale is determined by considering an excitation frequency $\omega$, corresponding to a wavelength of 500 nm. We note that the theoretical model does not encounter the physical sizes of the MNSs, hence does not encounter the retardation effects. Still, the time scale shown in Fig. \ref{fig:fig3} is consistent with the typical plasmon lifetime which is on the order of $1-100$ fs, in the visible spectrum. Here we demonstrate the comparative decaying behavior of plasmon energy, for the coupled and uncoupled cases. Red solid (dashed) line shows the normalized plasmon intensity in the coupled (uncoupled) cases. Absolute squares of the two coupled plasmon amplitudes are shown separately by black and blue lines. Inset graph shows the same, but zoomed in to the later times. For the uncoupled case, the number of plasmons in the system drops to zero at an earlier time. In the case of coupling, decay is slower, with the contribution of the attached oscillator (blue), which is excited by the driven oscillator.

\section{FDTD Simulations}

Next, we realize the problem that we treat with our theoretical model, by considering two physical systems, supporting bright and dark plasmon modes. The MNS that supports the bright plasmon mode is a Pt (platinum) nanosphere, whereas the one supporting the dark plasmon mode is an Au (gold) rectangular nanorod dimer. The nanorod dimer is composed of two identical nanorods, separated from each other along their axis, by a distance much smaller than their size. 

\begin{figure}
\includegraphics[width=\linewidth]{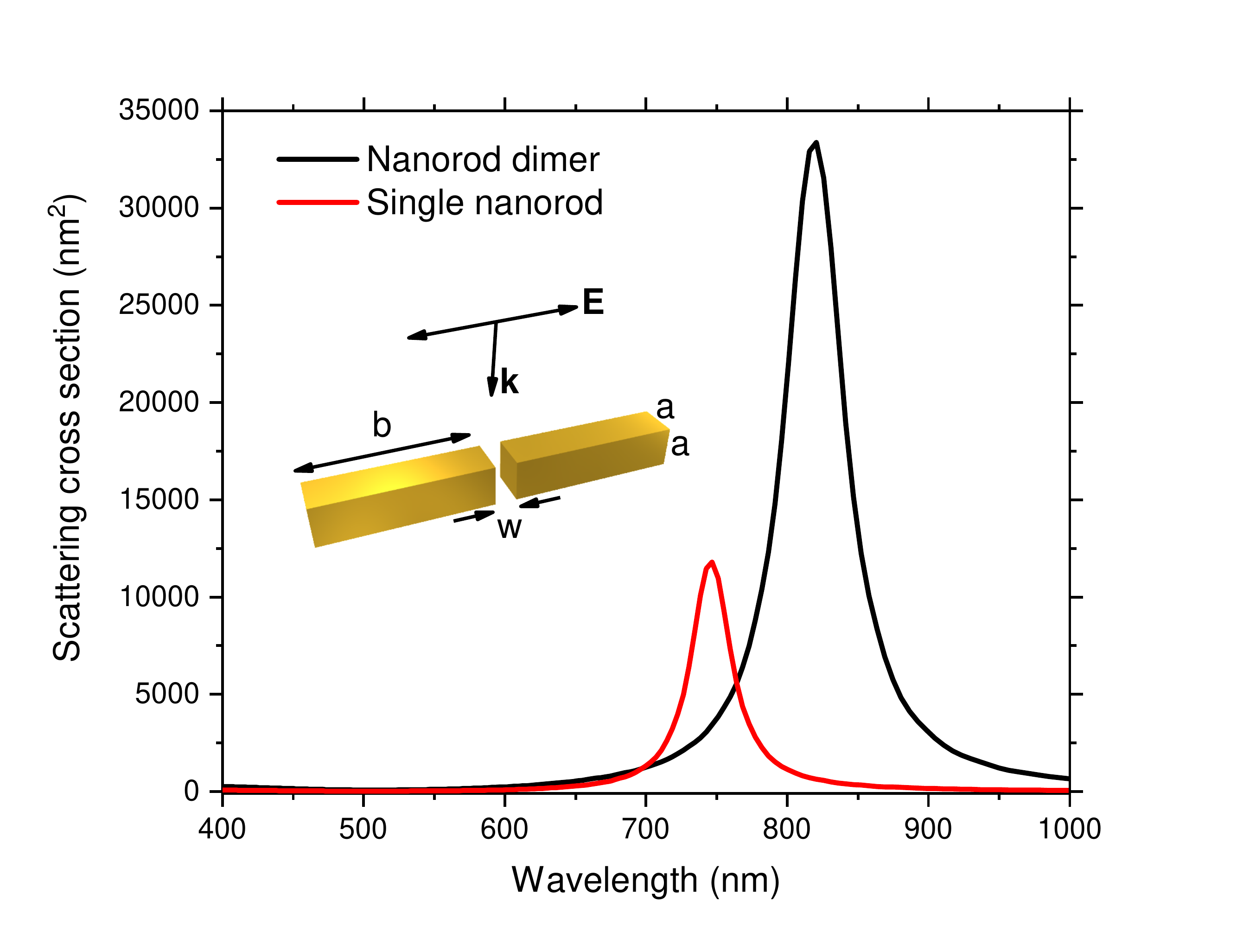}
\caption{\label{fig:fig4}Scattering cross section of a single nanorod and that of a nanorod dimer, obtained by FDTD simulations. The inset shows the schematics of the system with the given parameters:  $a=20$ nm, $b=80$ nm, $w=10$ nm. The incident plane wave is normally incident and polarized along the nanorod axis.}
\end{figure}

{\bf \small Characterization of the Au nanodimer}---
To understand the optical response of this MNS, we first calculate the scattering cross section of a single nanorod and that of the nanorod dimer. The light source is a plane wave illumination at normal incidence, polarized along the axis of the nanorod(s). We use the experimental dielectric function to model the Au, provided by Johnson and Christy \cite{Johnson1972}, and the surrounding medium is air. The boundaries of the simulation region are set to  perfectly matched layers (PMLs) which absorb all the light incident upon. The results are shown in Fig. \ref{fig:fig4}. The inset shows the schematics of the structure, where the given parameters are; $a=20$ nm, $b=80$ nm, $w=10$ nm. The resonance of the single nanorod is the excitation of longitudinal dipole plasmon mode \cite{C2CS35367A}, which is determined by the nanorod's aspect ratio. When two nanorods are brought together to form a dimer, the near-fields of the two nanorods interact with each other, resulting in split of the resonance into two hybrid resonances. Among those, the one observed in the scattering cross section of the nanorod dimer is the \textit{bonding} hybrid mode, having a net dipolar moment, and hence couples to the far-field, exhibiting a resonance. The other hybrid mode, however, possesses a vanishing dipole moment, due to symmetric charge alignments and hence cannot be excited by linearly polarized plane waves. This hybrid mode is called an \textit{anti-bonding}  dark mode, and can be excited by non-uniform electromagnetic field distributions, for example, a local dipole source. In what follows, we investigate the dark and bright hybrid plasmon modes of the gold nanorod dimer by employing a point dipole source.

\begin{figure}
\includegraphics[width=\linewidth]{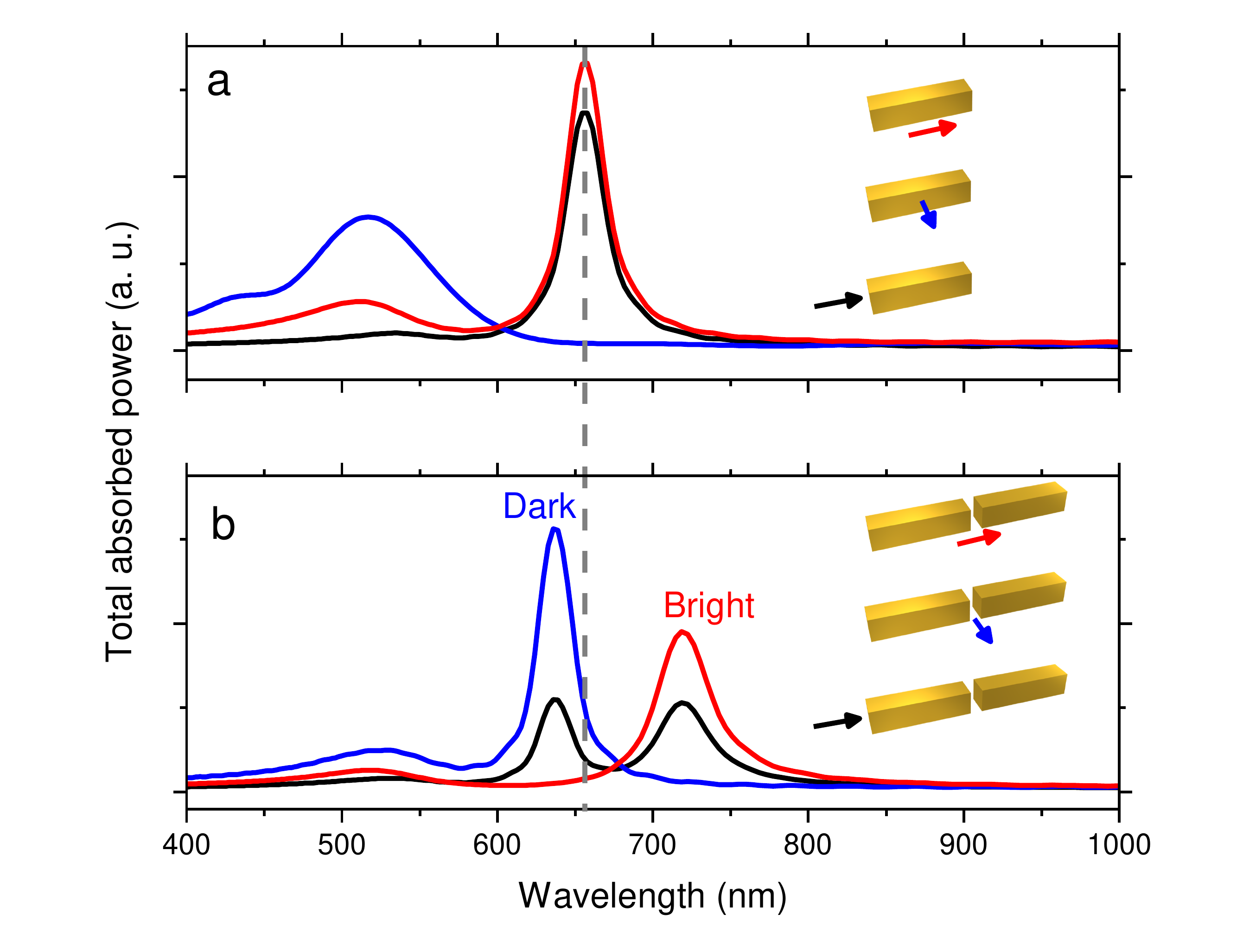}
\caption{\label{fig:fig5}Plasmon modes of a) a single nanorod, and b) a nanorod dimer, excited by local dipole sources; (i) red: in the middle, oriented along the nanorod axis, (ii) blue: in the middle, oriented perpendicular to the nanorod axis, and (iii) black: in the end, oriented along the nanorod axis. The distance between the dipole source and the nanostructure is 20 nm in each case, and the geometrical parameters are the same as the ones in Fig. \ref{fig:fig4}, i.e., nanorod length 80 nm, with a cross section of 20 nm x 20 nm and the separation between the nanorods in the dimer configuration is 10 nm.}
\end{figure}

We calculate the total absorbed power by the metallic nanostructures (single nanorod and nanorod dimer) where the nanostructures are excited with a dipole emitter source located at different positions and oriented at different directions, in FDTD simulations. The geometrical parameters and the dielectric properties are the same as the ones in the scattering cross section calculations. The dipole emitter source is positioned 20 nm away from the nanostructure (single nanorod or nanorod dimer) at three different manners: (i) in the middle, oriented along the nanorod axis, (ii) in the middle, oriented perpendicular to the nanorod axis, and (iii) in the end, oriented along the nanorod axis. These three different configurations are schematized in the insets of Fig. \ref{fig:fig5}, where the location and the orientation of the dipole emitters are illustrated by the arrows with different colors. The total absorbed power is calculated by box monitors enclosing the nanostructure, which measure the electric field intensity and electric permittivity. Figure \ref{fig:fig5}(a) shows the plasmonic modes of a single nanorod for three different dipole source configurations. When the dipole source is oriented along the nanorod axis, the plasmon mode is observed at $\sim660$ nm, which is the longitudinal mode. The longitudinal mode is excited more efficiently when the dipole source is located in the middle. The transverse plasmon mode of the nanorod, which is at $\sim515$ nm, is observed when the dipole emitter source is oriented perpendicular to the nanorod. Figure \ref{fig:fig5}(b) shows the plasmonic modes of the nanorod dimer. We observe that when the dipole source is at the end of the dimer, oriented perpendicular to the dimer axis, the absorption spectrum exhibits two plasmon modes, that are the hybrid modes, resulting from the coupling of the (individual) longitudinal plasmon modes of the two nanorods. The hybrid mode at the longer wavelength is the bright one, that is also observed in the scattering cross section spectrum presented in Fig. \ref{fig:fig4}. When the dipole emitter source, oriented along the dimer axis is in the middle of the dimer, only the bright hybrid mode of the nanorod dimer appears in the absorption spectrum. On the other hand, when a dipole source which is oriented perpendicular to the dimer axis is located in the middle, only the dark mode appears in the absorption spectrum. The reason why varying orientations and locations of the dipole source excite different modes is related with the coupling efficiency of the dipole moment, $\vec{p}$, of the dipole source and the local electric field $\vec{E}$ of the dimer, which is proportional to the inner product of these two vectors, $\vec{p} \cdotp \vec{E}$. 

\begin{figure}
\includegraphics[width=\linewidth]{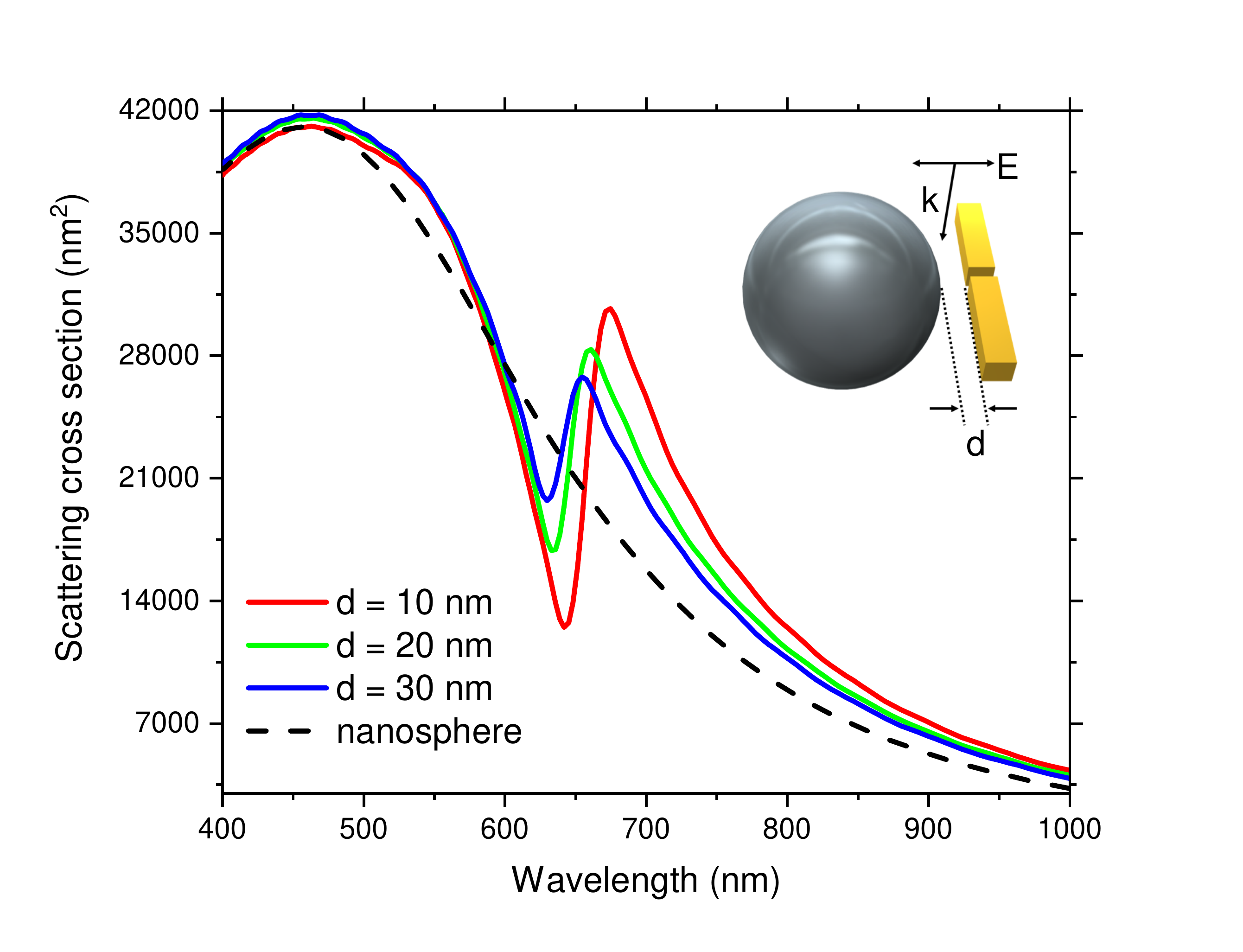}
\caption{\label{fig:fig6} Scattering cross section spectra of the Pt nanosphere of 75 nm radius (black dashed line) and the coupled systems with different inter-structure distances, $d$. The source is a plane wave at normal incidence, polarized perpendicular to the dimer axis, as shown in the inset. Nanorod length 80 nm, with a cross section of 20 nm x 20 nm and the separation between the nanorods is 10 nm.}
\end{figure}

{\bf \small The coupled system of Pt nanosphere and Au nanodimer}---
To study the effect of coupling between a dark and a bright plasmon mode to the decay properties of the system, we replace the local dipole emitter source with a plasmonic nanosphere having a dipolar plasmon mode, that is excited by a linearly polarized plane wave source. To avoid the complexities that would arise out of the presence of both bright and dark plasmon modes of the nanorod dimer, we allow only the dark mode to be excited by the near-field of the nanosphere, by setting the polarization of the incident field perpendicular to the dimer axis. In that sense, to excite (only) the dark mode of the dimer, we make use of the dipolar near-field of a metal nanosphere, which is excited by the plane wave in the desired orientation, rather than a local dipole source. We underline that the bright mode of the nanorod dimer is different than the one in the context of the theoretical model. The bright mode in the context of the theoretical model is the one supported by the Pt nanosphere, which is located in such a way that the overall structure looks like an antisymmetric trimer as illustrated in the inset of Fig. \ref{fig:fig6}.

We choose Pt for the metal nanosphere, as Pt nanostructures exhibit broader localized surface plasmon resonances compared to gold and and silver \cite{doi:10.1021/nl060219x}. It is an appropriate material to be studied as a short lifetime plasmonic oscillator in the context of the theoretical model. We use the experimental dielectric function provided by Palik to model Pt \cite{Palik1998HandbookSolids}. We calculate the scattering cross section of the Pt nanosphere, with a radius of 75 nm, and those of the coupled systems composed of the Pt nanosphere and the Au nanorod dimer, with inter-structure distances, $d=10, 20,$ and 30 nm. Figure \ref{fig:fig6} plots the results and the inset shows the geometrical configuration. We observe a quite broad plasmon resonance exhibited by the Pt nanosphere (black dashed line), as compared to that of the nanorod, and of the nanorod dimer shown in Fig. \ref{fig:fig4}. The resonance peaks at 500 nm and it covers a broad spectral wavelength range from ultraviolet to near-infrared. This resonance is modified by the presence of the nanorod dimer, in a way that it exhibits a sharp dip at $\sim 640$ nm, where the dark mode of the nanorod dimer is present. This indicates that the bright mode of the Pt nanosphere (excited by the incident field) and the dark mode of the Au dimer are coupled. Coupling between these bright and dark plasmon modes results in a narrowing in the resonance as evident from the scattering spectra \cite{and1999SpectralNanorods}. We note that the scattering cross section of the nanorod dimer in the absence of the the Pt nanosphere (not shown here) exhibits a peak at 450 nm, which is the transverse plasmon mode of the dimer. The presence of the transverse mode of the dimer may be the reason why the coupled resonance displays a broadening around 550 nm, compared to the resonance of the Pt nanoshpere, alone. So, the decay properties of the individual resonance of the Pt nanosphere are altered as a result of coupling to the dark plasmon mode.

\begin{figure}
\includegraphics[width=\linewidth]{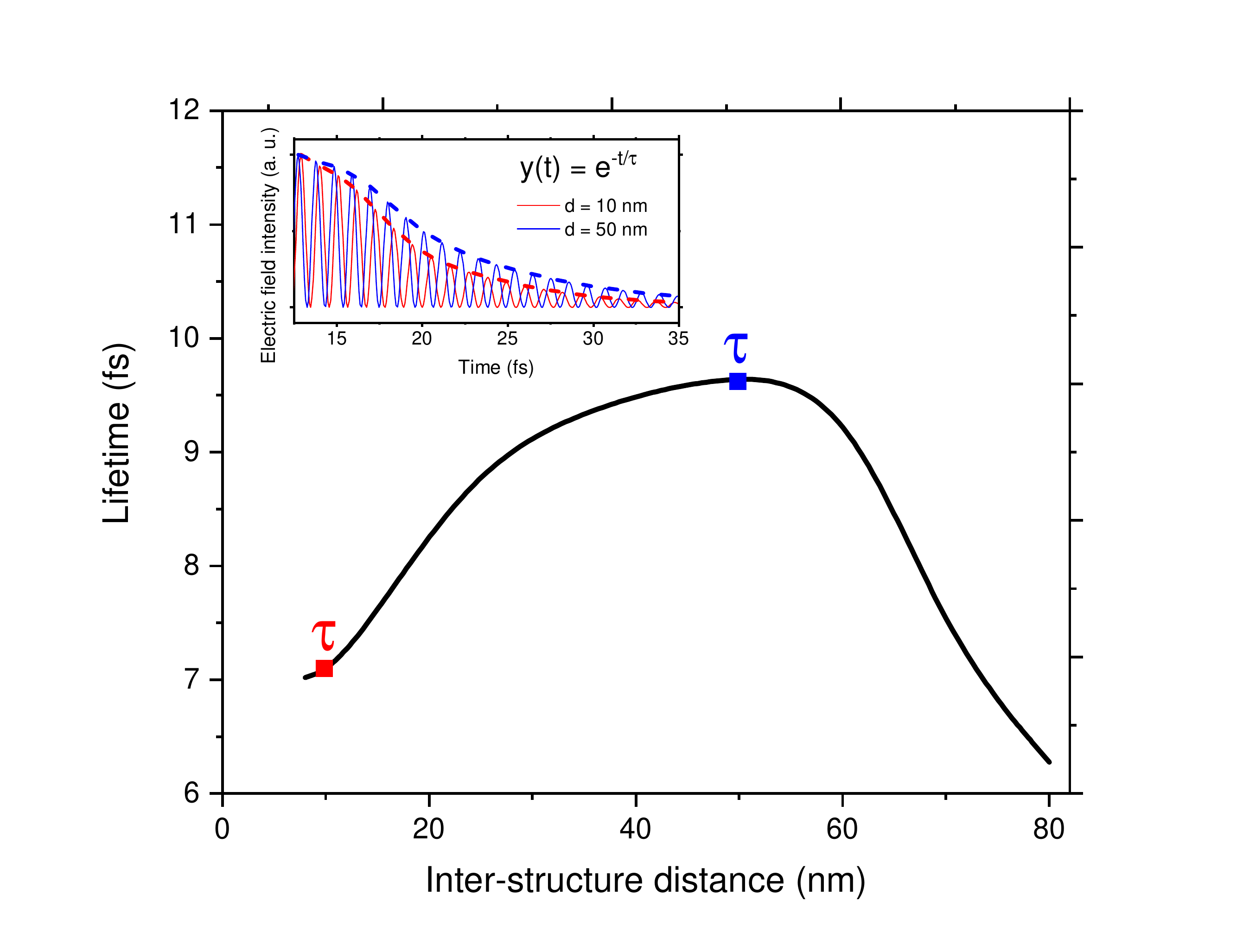}
\caption{\label{fig:fig7} Lifetime of the coupled system, defined by the exponential decay fit, that is applied to the electric field intensity over time, for different inter-structure distances. The inset shows the electric field intensity for $d = 10$ nm and $d = 50$ nm (solid lines), and the peaks of the oscillating field are shown by the dashed lines. The formula given in the inset is the fit formula to obtain $\tau$.}
\end{figure}

{\bf \small Hotspot lifetime enhancement}---
To get a better understanding of the modified decay properties of the coupled system, we investigate the decay of electromagnetic fields over time for changing inter-structure distance, so that we construct a link between our theoretical model's phenomenological coupling strength $f$, and the physical distance between the MNSs. We simulate how the electric fields decay in time for a range of inter-structure distance, by a point monitor located at the hotspot of the coupled system (at a distance $d/2$ from the nanosphere: see the illustration shown in Fig. \ref{fig:fig6}). We use a plane wave source normally incident on the coupled system, polarized perpendicular to the dimer axis, so that the longitudinal bright plasmon mode is not excited, but the dark plasmon mode is excited by the near-field of the nanosphere's plasmon mode. After we obtain the intensity of the electric field recorded by the monitor, as a function of time, we fit the simulation data to the exponential decay function, in the form of $I(t)=I_0e^{-t/\tau}$, where $\tau$ is defined as the lifetime of the oscillations, in the context of this study. Figure \ref{fig:fig7} shows the obtained results. The inset shows the electric field intensity, that is normalized to $I_0$, for $d = 10$ nm and $d = 50$ nm, by solid (oscillating) lines; and the dashed lines are the peak values of each field intensity, on which we apply the exponential fits. Corresponding lifetime ($\tau$) values obtained from the exponential fits are marked by the squares with associated colors on the lifetime curve. We observe that the lifetime of the system gets longer as the distance gets larger at first, reaches a maximum at particular distances, and then starts to drop again. When the inter-structure distance is larger than $\sim$ 150 nm, it drops down to 1.3 fs, which is the lifetime of the isolated nanosphere, the short-lifetime MNS. These confirm the outcomes of the theoretical model, where the coupling strength, $f$, gets an optimum, and the normalized lifetime reaches to 1 when the MNSs are no longer coupled, at $f=0$.
\section{Discussion and conclusions}

We examine the lifetime of a coupled plasmonic system of two MNSs, one of which can be excited by an incident field, referred as a bright plasmon mode. The plasmon mode of the excited MNS has a large damping rate, i.e. a short lifetime. The second MNS has a longer lifetime but cannot be excited by the external field, that is a dark plasmon mode. Bringing two such MNSs together, so that they interact with each other, we demonstrate that it is possible to modify the decay properties of the coupled system. Compared to the isolated MNS with shorter lifetime, coupled system decays more slowly for a wide range of inter-structure distance. 
We examine the coupled system based on two different approaches. In the first approach, we apply a simple analytical model based on the solutions of the Heisenberg equation, where the MNSs are treated as size-less (point-like) harmonic oscillators, and the coupling between them is quantified by a phenomenological constant. This approach, which may appear like oversimplified, yields outcomes, —in terms of the trend of the lifetime for varying coupling strengths— similar to the results obtained by 3D solutions of the Maxwell’s equations for finite-size MNSs. In the FDTD solutions, the retardation effects are fully accounted, and the coupling strength is quantified by the physical distance between the two MNSs. FDTD simulations show that the decay time of the electric field at the hotspot of the coupled system (at the gap between the Pt nanosphere and Au nanorod dimer) extends significantly, compared to the decay time of the nanosphere's near-field in the absence of the dimer. This significant enhancement is not observable in the scattering cross section calculations, i.e., far-field optical response. Because the scattering cross section measurements, conducted in experiments, do report the decay rate of the overall system, i.e. not the hotspot intensity. Distinguishing between the lifetime of the overall versus hotspot decays, could be useful for most of the imaging/detection and nonlinearity enhancement techniques. For instance, the signal (e.g. Rayleigh, fluorescence or Raman) in SNOM and SERS imaging is sensitive only to the events that take place in the hotspot, i.e., not the overall system. From the experimental point of view, the enhanced near-field lifetime could be observed by locating a reporter molecule at the hotspot of the coupled system, and measuring the off-resonant Raman scattering. Such a molecule would experience a rather different field in the time domain when it is at the hotspot, which is not expected, considering the slight cross section narrowing (considering the lifetime modification, extracted by the far-field observations, is not significant). Furthermore, FDTD simulations show that as the distance between the MNSs gets smaller, lifetime increases up to an inter-structure distance, and then it starts to decrease as the MNSs are brought even more closer, where strong hybridization starts. This behavior is similar to the observation of fluorescence enhancement near a MNS \cite{Anger2006EnhancementFluorescence}. Analogically, in the case of fluorescence enhancement, the fluorescent molecule is the long-lifetime object. 

We provide an extensive study on the lifetime of dark-bright plasmon modes, which would improve the current intuitive knowledge of the lifetime phenomenon. The results provide a deeper understanding of the classical mechanism of the coupling between two MNSs supporting bright and dark modes, depending on the distance from each other. The enhancement in the lifetime, obtained via coupling to a dark mode, offers implementations for technologies based on both linear and nonlinear response, e.g. solar cells and SERS, utilizing the plasmon localization. 

\section*{Acknowledgements}
M.E.T. acknowledges support from TUBITAK 1001-
117F118. A.B. and M.E.T. are supported by TUBITAK 1001-
119F101. M.E.T. gratefully thanks Halil Aydin and BMT
Calsis Health Technologies Co.

\bibliography{references}

\begin{thebibliography}{52}%
\makeatletter
\providecommand \@ifxundefined [1]{%
 \@ifx{#1\undefined}
}%
\providecommand \@ifnum [1]{%
 \ifnum #1\expandafter \@firstoftwo
 \else \expandafter \@secondoftwo
 \fi
}%
\providecommand \@ifx [1]{%
 \ifx #1\expandafter \@firstoftwo
 \else \expandafter \@secondoftwo
 \fi
}%
\providecommand \natexlab [1]{#1}%
\providecommand \enquote  [1]{``#1''}%
\providecommand \bibnamefont  [1]{#1}%
\providecommand \bibfnamefont [1]{#1}%
\providecommand \citenamefont [1]{#1}%
\providecommand \href@noop [0]{\@secondoftwo}%
\providecommand \href [0]{\begingroup \@sanitize@url \@href}%
\providecommand \@href[1]{\@@startlink{#1}\@@href}%
\providecommand \@@href[1]{\endgroup#1\@@endlink}%
\providecommand \@sanitize@url [0]{\catcode `\\12\catcode `\$12\catcode
  `\&12\catcode `\#12\catcode `\^12\catcode `\_12\catcode `\%12\relax}%
\providecommand \@@startlink[1]{}%
\providecommand \@@endlink[0]{}%
\providecommand \url  [0]{\begingroup\@sanitize@url \@url }%
\providecommand \@url [1]{\endgroup\@href {#1}{\urlprefix }}%
\providecommand \urlprefix  [0]{URL }%
\providecommand \Eprint [0]{\href }%
\providecommand \doibase [0]{https://doi.org/}%
\providecommand \selectlanguage [0]{\@gobble}%
\providecommand \bibinfo  [0]{\@secondoftwo}%
\providecommand \bibfield  [0]{\@secondoftwo}%
\providecommand \translation [1]{[#1]}%
\providecommand \BibitemOpen [0]{}%
\providecommand \bibitemStop [0]{}%
\providecommand \bibitemNoStop [0]{.\EOS\space}%
\providecommand \EOS [0]{\spacefactor3000\relax}%
\providecommand \BibitemShut  [1]{\csname bibitem#1\endcsname}%
\let\auto@bib@innerbib\@empty
\bibitem [{\citenamefont {Stockman}(2011)}]{Stockman2011Nanoplasmonics:Future}%
  \BibitemOpen
  \bibfield  {author} {\bibinfo {author} {\bibfnamefont {M.~I.}\ \bibnamefont
  {Stockman}},\ }\bibfield  {title} {\bibinfo {title} {{Nanoplasmonics: past,
  present, and glimpse into future}},\ }\href
  {https://doi.org/10.1364/OE.19.022029} {\bibfield  {journal} {\bibinfo
  {journal} {Optics Express}\ }\textbf {\bibinfo {volume} {19}},\ \bibinfo
  {pages} {22029} (\bibinfo {year} {2011})}\BibitemShut {NoStop}%
\bibitem [{\citenamefont {Anker}\ \emph {et~al.}(2008)\citenamefont {Anker},
  \citenamefont {Hall}, \citenamefont {Lyandres}, \citenamefont {Shah},
  \citenamefont {Zhao},\ and\ \citenamefont
  {Van~Duyne}}]{Anker2008BiosensingNanosensors}%
  \BibitemOpen
  \bibfield  {author} {\bibinfo {author} {\bibfnamefont {J.~N.}\ \bibnamefont
  {Anker}}, \bibinfo {author} {\bibfnamefont {W.~P.}\ \bibnamefont {Hall}},
  \bibinfo {author} {\bibfnamefont {O.}~\bibnamefont {Lyandres}}, \bibinfo
  {author} {\bibfnamefont {N.~C.}\ \bibnamefont {Shah}}, \bibinfo {author}
  {\bibfnamefont {J.}~\bibnamefont {Zhao}}, and\ \bibinfo {author}
  {\bibfnamefont {R.~P.}\ \bibnamefont {Van~Duyne}},\ }\bibfield  {title}
  {\bibinfo {title} {{Biosensing with plasmonic nanosensors}},\ }\href
  {https://doi.org/10.1038/nmat2162} {\bibfield  {journal} {\bibinfo  {journal}
  {Nature Materials}\ }\textbf {\bibinfo {volume} {7}},\ \bibinfo {pages} {442}
  (\bibinfo {year} {2008})}\BibitemShut {NoStop}%
\bibitem [{\citenamefont {Liu}\ \emph {et~al.}(2015)\citenamefont {Liu},
  \citenamefont {Liu}, \citenamefont {Chen}, \citenamefont {Cheng},
  \citenamefont {Wang},\ and\ \citenamefont {Peng}}]{Liu2015SurfacePlatforms}%
  \BibitemOpen
  \bibfield  {author} {\bibinfo {author} {\bibfnamefont {Y.}~\bibnamefont
  {Liu}}, \bibinfo {author} {\bibfnamefont {Q.}~\bibnamefont {Liu}}, \bibinfo
  {author} {\bibfnamefont {S.}~\bibnamefont {Chen}}, \bibinfo {author}
  {\bibfnamefont {F.}~\bibnamefont {Cheng}}, \bibinfo {author} {\bibfnamefont
  {H.}~\bibnamefont {Wang}}, and\ \bibinfo {author} {\bibfnamefont
  {W.}~\bibnamefont {Peng}},\ }\bibfield  {title} {\bibinfo {title} {{Surface
  Plasmon Resonance Biosensor Based on Smart Phone Platforms}},\ }\href
  {https://doi.org/10.1038/srep12864} {\bibfield  {journal} {\bibinfo
  {journal} {Scientific Reports}\ }\textbf {\bibinfo {volume} {5}},\ \bibinfo
  {pages} {12864} (\bibinfo {year} {2015})}\BibitemShut {NoStop}%
\bibitem [{\citenamefont {Mej{\'{i}}a-Salazar}\ and\ \citenamefont
  {Oliveira}(2018)}]{Mejia-Salazar2018PlasmonicBiosensing}%
  \BibitemOpen
  \bibfield  {author} {\bibinfo {author} {\bibfnamefont {J.~R.}\ \bibnamefont
  {Mej{\'{i}}a-Salazar}}and\ \bibinfo {author} {\bibfnamefont {O.~N.}\
  \bibnamefont {Oliveira}},\ }\bibfield  {title} {\bibinfo {title} {{Plasmonic
  Biosensing}},\ }\href {https://doi.org/10.1021/acs.chemrev.8b00359}
  {\bibfield  {journal} {\bibinfo  {journal} {Chemical Reviews}\ }\textbf
  {\bibinfo {volume} {118}},\ \bibinfo {pages} {10617} (\bibinfo {year}
  {2018})}\BibitemShut {NoStop}%
\bibitem [{\citenamefont {Dregely}\ \emph {et~al.}(2011)\citenamefont
  {Dregely}, \citenamefont {Taubert}, \citenamefont {Dorfm{\"{u}}ller},
  \citenamefont {Vogelgesang}, \citenamefont {Kern},\ and\ \citenamefont
  {Giessen}}]{Dregely20113DArray}%
  \BibitemOpen
  \bibfield  {author} {\bibinfo {author} {\bibfnamefont {D.}~\bibnamefont
  {Dregely}}, \bibinfo {author} {\bibfnamefont {R.}~\bibnamefont {Taubert}},
  \bibinfo {author} {\bibfnamefont {J.}~\bibnamefont {Dorfm{\"{u}}ller}},
  \bibinfo {author} {\bibfnamefont {R.}~\bibnamefont {Vogelgesang}}, \bibinfo
  {author} {\bibfnamefont {K.}~\bibnamefont {Kern}}, and\ \bibinfo {author}
  {\bibfnamefont {H.}~\bibnamefont {Giessen}},\ }\bibfield  {title} {\bibinfo
  {title} {{3D optical Yagi–Uda nanoantenna array}},\ }\href
  {https://doi.org/10.1038/ncomms1268} {\bibfield  {journal} {\bibinfo
  {journal} {Nature Communications}\ }\textbf {\bibinfo {volume} {2}},\
  \bibinfo {pages} {267} (\bibinfo {year} {2011})}\BibitemShut {NoStop}%
\bibitem [{\citenamefont {Akselrod}\ \emph {et~al.}(2014)\citenamefont
  {Akselrod}, \citenamefont {Argyropoulos}, \citenamefont {Hoang},
  \citenamefont {Cirac{\`{i}}}, \citenamefont {Fang}, \citenamefont {Huang},
  \citenamefont {Smith},\ and\ \citenamefont
  {Mikkelsen}}]{Akselrod2014ProbingNanoantennas}%
  \BibitemOpen
  \bibfield  {author} {\bibinfo {author} {\bibfnamefont {G.~M.}\ \bibnamefont
  {Akselrod}}, \bibinfo {author} {\bibfnamefont {C.}~\bibnamefont
  {Argyropoulos}}, \bibinfo {author} {\bibfnamefont {T.~B.}\ \bibnamefont
  {Hoang}}, \bibinfo {author} {\bibfnamefont {C.}~\bibnamefont {Cirac{\`{i}}}},
  \bibinfo {author} {\bibfnamefont {C.}~\bibnamefont {Fang}}, \bibinfo {author}
  {\bibfnamefont {J.}~\bibnamefont {Huang}}, \bibinfo {author} {\bibfnamefont
  {D.~R.}\ \bibnamefont {Smith}}, and\ \bibinfo {author} {\bibfnamefont
  {M.~H.}\ \bibnamefont {Mikkelsen}},\ }\bibfield  {title} {\bibinfo {title}
  {{Probing the mechanisms of large Purcell enhancement in plasmonic
  nanoantennas}},\ }\href {https://doi.org/10.1038/nphoton.2014.228} {\bibfield
   {journal} {\bibinfo  {journal} {Nature Photonics}\ }\textbf {\bibinfo
  {volume} {8}},\ \bibinfo {pages} {835} (\bibinfo {year} {2014})}\BibitemShut
  {NoStop}%
\bibitem [{\citenamefont {Savaliya}\ \emph {et~al.}(2017)\citenamefont
  {Savaliya}, \citenamefont {Thomas}, \citenamefont {Dua},\ and\ \citenamefont
  {Dhawan}}]{Savaliya2017TunableMaterials}%
  \BibitemOpen
  \bibfield  {author} {\bibinfo {author} {\bibfnamefont {P.~B.}\ \bibnamefont
  {Savaliya}}, \bibinfo {author} {\bibfnamefont {A.}~\bibnamefont {Thomas}},
  \bibinfo {author} {\bibfnamefont {R.}~\bibnamefont {Dua}}, and\ \bibinfo
  {author} {\bibfnamefont {A.}~\bibnamefont {Dhawan}},\ }\bibfield  {title}
  {\bibinfo {title} {{Tunable optical switching in the near-infrared spectral
  regime by employing plasmonic nanoantennas containing phase change
  materials}},\ }\href {https://doi.org/10.1364/OE.25.023755} {\bibfield
  {journal} {\bibinfo  {journal} {Optics Express}\ }\textbf {\bibinfo {volume}
  {25}},\ \bibinfo {pages} {23755} (\bibinfo {year} {2017})}\BibitemShut
  {NoStop}%
\bibitem [{\citenamefont {Yildiz}\ \emph {et~al.}(2019)\citenamefont {Yildiz},
  \citenamefont {Habib}, \citenamefont {Rashed},\ and\ \citenamefont
  {Caglayan}}]{Yildiz2019}%
  \BibitemOpen
  \bibfield  {author} {\bibinfo {author} {\bibfnamefont {B.~C.}\ \bibnamefont
  {Yildiz}}, \bibinfo {author} {\bibfnamefont {M.}~\bibnamefont {Habib}},
  \bibinfo {author} {\bibfnamefont {A.~R.}\ \bibnamefont {Rashed}}, and\
  \bibinfo {author} {\bibfnamefont {H.}~\bibnamefont {Caglayan}},\ }\bibfield
  {title} {\bibinfo {title} {Hybridized plasmon modes in a system of metal thin
  film - nanodisk array},\ }\href {https://doi.org/10.1063/1.5115818}
  {\bibfield  {journal} {\bibinfo  {journal} {Journal of Applied Physics}\
  }\textbf {\bibinfo {volume} {126}},\ \bibinfo {pages} {113104} (\bibinfo
  {year} {2019})}\BibitemShut {NoStop}%
\bibitem [{\citenamefont {Fang}\ \emph {et~al.}(2005)\citenamefont {Fang},
  \citenamefont {Lee}, \citenamefont {Sun},\ and\ \citenamefont
  {Zhang}}]{Fang2005Sub-Diffraction-LimitedSuperlens}%
  \BibitemOpen
  \bibfield  {author} {\bibinfo {author} {\bibfnamefont {N.}~\bibnamefont
  {Fang}}, \bibinfo {author} {\bibfnamefont {H.}~\bibnamefont {Lee}}, \bibinfo
  {author} {\bibfnamefont {C.}~\bibnamefont {Sun}}, and\ \bibinfo {author}
  {\bibfnamefont {X.}~\bibnamefont {Zhang}},\ }\bibfield  {title} {\bibinfo
  {title} {{Sub-Diffraction-Limited Optical Imaging with a Silver Superlens}},\
  }\href {https://doi.org/10.1126/science.1108759} {\bibfield  {journal}
  {\bibinfo  {journal} {Science}\ }\textbf {\bibinfo {volume} {308}},\ \bibinfo
  {pages} {534} (\bibinfo {year} {2005})}\BibitemShut {NoStop}%
\bibitem [{\citenamefont {Maier}(2007)}]{Maier2007l}%
  \BibitemOpen
  \bibfield  {author} {\bibinfo {author} {\bibfnamefont {S.~A.}\ \bibnamefont
  {Maier}},\ }\bibfield  {title} {\bibinfo {title} {{Metamaterials and Imaging
  with Surface Plasmon Polaritons}},\ }in\ \href
  {https://doi.org/10.1007/0-387-37825-1{\_}11} {\emph {\bibinfo {booktitle}
  {Plasmonics: Fundamentals and Applications}}}\ (\bibinfo  {publisher}
  {Springer US},\ \bibinfo {address} {New York, NY},\ \bibinfo {year} {2007})\
  pp.\ \bibinfo {pages} {193--200}\BibitemShut {NoStop}%
\bibitem [{\citenamefont {Kawata}\ \emph {et~al.}(2009)\citenamefont {Kawata},
  \citenamefont {Inouye},\ and\ \citenamefont {Verma}}]{Kawata2009}%
  \BibitemOpen
  \bibfield  {author} {\bibinfo {author} {\bibfnamefont {S.}~\bibnamefont
  {Kawata}}, \bibinfo {author} {\bibfnamefont {Y.}~\bibnamefont {Inouye}}, and\
  \bibinfo {author} {\bibfnamefont {P.}~\bibnamefont {Verma}},\ }\bibfield
  {title} {\bibinfo {title} {{Plasmonics for near-field nano-imaging and
  superlensing}},\ }\href {https://doi.org/10.1038/nphoton.2009.111} {\bibfield
   {journal} {\bibinfo  {journal} {Nature Photonics}\ }\textbf {\bibinfo
  {volume} {3}},\ \bibinfo {pages} {388} (\bibinfo {year} {2009})}\BibitemShut
  {NoStop}%
\bibitem [{\citenamefont {Pompa}\ \emph {et~al.}(2006)\citenamefont {Pompa},
  \citenamefont {Martiradonna}, \citenamefont {Torre}, \citenamefont {Sala},
  \citenamefont {Manna}, \citenamefont {De~Vittorio}, \citenamefont {Calabi},
  \citenamefont {Cingolani},\ and\ \citenamefont {Rinaldi}}]{Pompa2006}%
  \BibitemOpen
  \bibfield  {author} {\bibinfo {author} {\bibfnamefont {P.~P.}\ \bibnamefont
  {Pompa}}, \bibinfo {author} {\bibfnamefont {L.}~\bibnamefont {Martiradonna}},
  \bibinfo {author} {\bibfnamefont {A.~D.}\ \bibnamefont {Torre}}, \bibinfo
  {author} {\bibfnamefont {F.~D.}\ \bibnamefont {Sala}}, \bibinfo {author}
  {\bibfnamefont {L.}~\bibnamefont {Manna}}, \bibinfo {author} {\bibfnamefont
  {M.}~\bibnamefont {De~Vittorio}}, \bibinfo {author} {\bibfnamefont
  {F.}~\bibnamefont {Calabi}}, \bibinfo {author} {\bibfnamefont
  {R.}~\bibnamefont {Cingolani}}, and\ \bibinfo {author} {\bibfnamefont
  {R.}~\bibnamefont {Rinaldi}},\ }\bibfield  {title} {\bibinfo {title}
  {{Metal-enhanced fluorescence of colloidal nanocrystals with nanoscale
  control}},\ }\href {https://doi.org/10.1038/nnano.2006.93} {\bibfield
  {journal} {\bibinfo  {journal} {Nature Nanotechnology}\ }\textbf {\bibinfo
  {volume} {1}},\ \bibinfo {pages} {126} (\bibinfo {year} {2006})}\BibitemShut
  {NoStop}%
\bibitem [{\citenamefont {Bharadwaj}\ and\ \citenamefont
  {Novotny}(2007)}]{Bharadwaj2007SpectralEnhancement}%
  \BibitemOpen
  \bibfield  {author} {\bibinfo {author} {\bibfnamefont {P.}~\bibnamefont
  {Bharadwaj}}and\ \bibinfo {author} {\bibfnamefont {L.}~\bibnamefont
  {Novotny}},\ }\bibfield  {title} {\bibinfo {title} {{Spectral dependence of
  single molecule fluorescence enhancement}},\ }\href
  {https://doi.org/10.1364/OE.15.014266} {\bibfield  {journal} {\bibinfo
  {journal} {Optics Express}\ }\textbf {\bibinfo {volume} {15}},\ \bibinfo
  {pages} {14266} (\bibinfo {year} {2007})}\BibitemShut {NoStop}%
\bibitem [{\citenamefont {Liu}\ \emph {et~al.}(2013)\citenamefont {Liu},
  \citenamefont {Huang}, \citenamefont {Li}, \citenamefont {Wang},
  \citenamefont {Li}, \citenamefont {Xu}, \citenamefont {Guo}, \citenamefont
  {Meng}, \citenamefont {Shi},\ and\ \citenamefont {Li}}]{Liu2013}%
  \BibitemOpen
  \bibfield  {author} {\bibinfo {author} {\bibfnamefont {S.~Y.}\ \bibnamefont
  {Liu}}, \bibinfo {author} {\bibfnamefont {L.}~\bibnamefont {Huang}}, \bibinfo
  {author} {\bibfnamefont {J.~F.}\ \bibnamefont {Li}}, \bibinfo {author}
  {\bibfnamefont {C.}~\bibnamefont {Wang}}, \bibinfo {author} {\bibfnamefont
  {Q.}~\bibnamefont {Li}}, \bibinfo {author} {\bibfnamefont {H.~X.}\
  \bibnamefont {Xu}}, \bibinfo {author} {\bibfnamefont {H.~L.}\ \bibnamefont
  {Guo}}, \bibinfo {author} {\bibfnamefont {Z.~M.}\ \bibnamefont {Meng}},
  \bibinfo {author} {\bibfnamefont {Z.}~\bibnamefont {Shi}}, and\ \bibinfo
  {author} {\bibfnamefont {Z.~Y.}\ \bibnamefont {Li}},\ }\bibfield  {title}
  {\bibinfo {title} {{Simultaneous excitation and emission enhancement of
  fluorescence assisted by double plasmon modes of gold nanorods}},\ }\href
  {https://doi.org/10.1021/jp4001626} {\bibfield  {journal} {\bibinfo
  {journal} {Journal of Physical Chemistry C}\ }\textbf {\bibinfo {volume}
  {117}},\ \bibinfo {pages} {10636} (\bibinfo {year} {2013})}\BibitemShut
  {NoStop}%
\bibitem [{\citenamefont {Carre{\~{n}}o}\ \emph {et~al.}(2016)\citenamefont
  {Carre{\~{n}}o}, \citenamefont {Ant{\'{o}}n}, \citenamefont {Yannopapas},\
  and\ \citenamefont {Paspalakis}}]{Carreno2016ResonanceNanoparticle}%
  \BibitemOpen
  \bibfield  {author} {\bibinfo {author} {\bibfnamefont {F.}~\bibnamefont
  {Carre{\~{n}}o}}, \bibinfo {author} {\bibfnamefont {M.~A.}\ \bibnamefont
  {Ant{\'{o}}n}}, \bibinfo {author} {\bibfnamefont {V.}~\bibnamefont
  {Yannopapas}}, and\ \bibinfo {author} {\bibfnamefont {E.}~\bibnamefont
  {Paspalakis}},\ }\bibfield  {title} {\bibinfo {title} {{Resonance
  fluorescence spectrum of a {$\Lambda$} -type quantum emitter close to a
  metallic nanoparticle}},\ }\href {https://doi.org/10.1103/PhysRevA.94.013834}
  {\bibfield  {journal} {\bibinfo  {journal} {Physical Review A}\ }\textbf
  {\bibinfo {volume} {94}},\ \bibinfo {pages} {013834} (\bibinfo {year}
  {2016})}\BibitemShut {NoStop}%
\bibitem [{\citenamefont {Hsu}\ \emph {et~al.}(2017)\citenamefont {Hsu},
  \citenamefont {Ding},\ and\ \citenamefont
  {Schatz}}]{Hsu2017Plasmon-CoupledTransfer}%
  \BibitemOpen
  \bibfield  {author} {\bibinfo {author} {\bibfnamefont {L.-Y.}\ \bibnamefont
  {Hsu}}, \bibinfo {author} {\bibfnamefont {W.}~\bibnamefont {Ding}}, and\
  \bibinfo {author} {\bibfnamefont {G.~C.}\ \bibnamefont {Schatz}},\ }\bibfield
   {title} {\bibinfo {title} {{Plasmon-Coupled Resonance Energy Transfer}},\
  }\href {https://doi.org/10.1021/acs.jpclett.7b00526} {\bibfield  {journal}
  {\bibinfo  {journal} {The Journal of Physical Chemistry Letters}\ }\textbf
  {\bibinfo {volume} {8}},\ \bibinfo {pages} {2357} (\bibinfo {year}
  {2017})}\BibitemShut {NoStop}%
\bibitem [{\citenamefont {Garcia-Vidal}\ \emph {et~al.}(2005)\citenamefont
  {Garcia-Vidal}, \citenamefont {Mart{\'{i}}n-Moreno},\ and\ \citenamefont
  {Pendry}}]{Garcia-Vidal2005}%
  \BibitemOpen
  \bibfield  {author} {\bibinfo {author} {\bibfnamefont {F.~J.}\ \bibnamefont
  {Garcia-Vidal}}, \bibinfo {author} {\bibfnamefont {L.}~\bibnamefont
  {Mart{\'{i}}n-Moreno}}, and\ \bibinfo {author} {\bibfnamefont {J.~B.}\
  \bibnamefont {Pendry}},\ }\bibfield  {title} {\bibinfo {title} {{Surfaces
  with holes in them: new plasmonic metamaterials}},\ }\href
  {https://doi.org/10.1088/1464-4258/7/2/013} {\bibfield  {journal} {\bibinfo
  {journal} {Journal of Optics A: Pure and Applied Optics}\ }\textbf {\bibinfo
  {volume} {7}},\ \bibinfo {pages} {S97} (\bibinfo {year} {2005})}\BibitemShut
  {NoStop}%
\bibitem [{\citenamefont {Zheludev}(2011)}]{Zheludev2011}%
  \BibitemOpen
  \bibfield  {author} {\bibinfo {author} {\bibfnamefont {N.~I.}\ \bibnamefont
  {Zheludev}},\ }\bibfield  {title} {\bibinfo {title} {{A Roadmap for
  Metamaterials}},\ }\href {https://doi.org/10.1364/OPN.22.3.000030} {\bibfield
   {journal} {\bibinfo  {journal} {Optics and Photonics News}\ }\textbf
  {\bibinfo {volume} {22}},\ \bibinfo {pages} {30} (\bibinfo {year}
  {2011})}\BibitemShut {NoStop}%
\bibitem [{\citenamefont {Naik}\ \emph {et~al.}(2013)\citenamefont {Naik},
  \citenamefont {Shalaev},\ and\ \citenamefont {Boltasseva}}]{Naik2013}%
  \BibitemOpen
  \bibfield  {author} {\bibinfo {author} {\bibfnamefont {G.~V.}\ \bibnamefont
  {Naik}}, \bibinfo {author} {\bibfnamefont {V.~M.}\ \bibnamefont {Shalaev}},
  and\ \bibinfo {author} {\bibfnamefont {A.}~\bibnamefont {Boltasseva}},\
  }\bibfield  {title} {\bibinfo {title} {{Alternative Plasmonic Materials:
  Beyond Gold and Silver}},\ }\href {https://doi.org/10.1002/adma.201205076}
  {\bibfield  {journal} {\bibinfo  {journal} {Advanced Materials}\ }\textbf
  {\bibinfo {volume} {25}},\ \bibinfo {pages} {3264} (\bibinfo {year}
  {2013})}\BibitemShut {NoStop}%
\bibitem [{\citenamefont {Kawata}(2013)}]{Kawata2013}%
  \BibitemOpen
  \bibfield  {author} {\bibinfo {author} {\bibfnamefont {S.}~\bibnamefont
  {Kawata}},\ }\bibfield  {title} {\bibinfo {title} {{Plasmonics for
  Nanoimaging and Nanospectroscopy}},\ }\href
  {https://doi.org/10.1366/12-06861} {\bibfield  {journal} {\bibinfo  {journal}
  {Applied Spectroscopy}\ }\textbf {\bibinfo {volume} {67}},\ \bibinfo {pages}
  {117} (\bibinfo {year} {2013})}\BibitemShut {NoStop}%
\bibitem [{\citenamefont {Zhang}\ \emph {et~al.}(2015)\citenamefont {Zhang},
  \citenamefont {Li}, \citenamefont {Tang}, \citenamefont {Fang}, \citenamefont
  {Wang}, \citenamefont {Huang}, \citenamefont {Liu}, \citenamefont {Zheng},
  \citenamefont {Cui},\ and\ \citenamefont {Mei}}]{Zhang2015}%
  \BibitemOpen
  \bibfield  {author} {\bibinfo {author} {\bibfnamefont {J.}~\bibnamefont
  {Zhang}}, \bibinfo {author} {\bibfnamefont {J.}~\bibnamefont {Li}}, \bibinfo
  {author} {\bibfnamefont {S.}~\bibnamefont {Tang}}, \bibinfo {author}
  {\bibfnamefont {Y.}~\bibnamefont {Fang}}, \bibinfo {author} {\bibfnamefont
  {J.}~\bibnamefont {Wang}}, \bibinfo {author} {\bibfnamefont {G.}~\bibnamefont
  {Huang}}, \bibinfo {author} {\bibfnamefont {R.}~\bibnamefont {Liu}}, \bibinfo
  {author} {\bibfnamefont {L.}~\bibnamefont {Zheng}}, \bibinfo {author}
  {\bibfnamefont {X.}~\bibnamefont {Cui}}, and\ \bibinfo {author}
  {\bibfnamefont {Y.}~\bibnamefont {Mei}},\ }\bibfield  {title} {\bibinfo
  {title} {{Whispering-gallery nanocavity plasmon-enhanced Raman
  spectroscopy}},\ }\href {https://doi.org/10.1038/srep15012} {\bibfield
  {journal} {\bibinfo  {journal} {Scientific Reports}\ }\textbf {\bibinfo
  {volume} {5}},\ \bibinfo {pages} {1} (\bibinfo {year} {2015})}\BibitemShut
  {NoStop}%
\bibitem [{\citenamefont {Yildiz}\ \emph {et~al.}(2015)\citenamefont {Yildiz},
  \citenamefont {Tasgin}, \citenamefont {Abak}, \citenamefont {Coskun},
  \citenamefont {Unalan},\ and\ \citenamefont
  {Bek}}]{Yildiz2015EnhancedNanostructures}%
  \BibitemOpen
  \bibfield  {author} {\bibinfo {author} {\bibfnamefont {B.~C.}\ \bibnamefont
  {Yildiz}}, \bibinfo {author} {\bibfnamefont {M.~E.}\ \bibnamefont {Tasgin}},
  \bibinfo {author} {\bibfnamefont {M.~K.}\ \bibnamefont {Abak}}, \bibinfo
  {author} {\bibfnamefont {S.}~\bibnamefont {Coskun}}, \bibinfo {author}
  {\bibfnamefont {H.~E.}\ \bibnamefont {Unalan}}, and\ \bibinfo {author}
  {\bibfnamefont {A.}~\bibnamefont {Bek}},\ }\bibfield  {title} {\bibinfo
  {title} {{Enhanced second harmonic generation from coupled asymmetric
  plasmonic metal nanostructures}},\ }\href
  {https://doi.org/10.1088/2040-8978/17/12/125005} {\bibfield  {journal}
  {\bibinfo  {journal} {Journal of Optics}\ }\textbf {\bibinfo {volume} {17}},\
  \bibinfo {pages} {125005} (\bibinfo {year} {2015})}\BibitemShut {NoStop}%
\bibitem [{\citenamefont {Singh}\ \emph {et~al.}(2016)\citenamefont {Singh},
  \citenamefont {Abak},\ and\ \citenamefont
  {Tasgin}}]{Singh2016EnhancementPaths}%
  \BibitemOpen
  \bibfield  {author} {\bibinfo {author} {\bibfnamefont {S.~K.}\ \bibnamefont
  {Singh}}, \bibinfo {author} {\bibfnamefont {M.~K.}\ \bibnamefont {Abak}},
  and\ \bibinfo {author} {\bibfnamefont {M.~E.}\ \bibnamefont {Tasgin}},\
  }\bibfield  {title} {\bibinfo {title} {{Enhancement of four-wave mixing via
  interference of multiple plasmonic conversion paths}},\ }\href
  {https://doi.org/10.1103/PhysRevB.93.035410} {\bibfield  {journal} {\bibinfo
  {journal} {Physical Review B}\ }\textbf {\bibinfo {volume} {93}},\ \bibinfo
  {pages} {035410} (\bibinfo {year} {2016})}\BibitemShut {NoStop}%
\bibitem [{\citenamefont {Drachev}\ \emph {et~al.}(2018)\citenamefont
  {Drachev}, \citenamefont {Kildishev}, \citenamefont {Borneman}, \citenamefont
  {Chen}, \citenamefont {Shalaev}, \citenamefont {Yamnitskiy}, \citenamefont
  {Norwood}, \citenamefont {Peyghambarian}, \citenamefont {Marder},
  \citenamefont {Padilha}, \citenamefont {Webster}, \citenamefont {Ensley},
  \citenamefont {Hagan},\ and\ \citenamefont
  {Van~Stryland}}]{Drachev2018EngineeredArray}%
  \BibitemOpen
  \bibfield  {author} {\bibinfo {author} {\bibfnamefont {V.~P.}\ \bibnamefont
  {Drachev}}, \bibinfo {author} {\bibfnamefont {A.~V.}\ \bibnamefont
  {Kildishev}}, \bibinfo {author} {\bibfnamefont {J.~D.}\ \bibnamefont
  {Borneman}}, \bibinfo {author} {\bibfnamefont {K.-P.}\ \bibnamefont {Chen}},
  \bibinfo {author} {\bibfnamefont {V.~M.}\ \bibnamefont {Shalaev}}, \bibinfo
  {author} {\bibfnamefont {K.}~\bibnamefont {Yamnitskiy}}, \bibinfo {author}
  {\bibfnamefont {R.~A.}\ \bibnamefont {Norwood}}, \bibinfo {author}
  {\bibfnamefont {N.}~\bibnamefont {Peyghambarian}}, \bibinfo {author}
  {\bibfnamefont {S.~R.}\ \bibnamefont {Marder}}, \bibinfo {author}
  {\bibfnamefont {L.~A.}\ \bibnamefont {Padilha}}, \bibinfo {author}
  {\bibfnamefont {S.}~\bibnamefont {Webster}}, \bibinfo {author} {\bibfnamefont
  {T.~R.}\ \bibnamefont {Ensley}}, \bibinfo {author} {\bibfnamefont {D.~J.}\
  \bibnamefont {Hagan}}, and\ \bibinfo {author} {\bibfnamefont {E.~W.}\
  \bibnamefont {Van~Stryland}},\ }\bibfield  {title} {\bibinfo {title}
  {{Engineered nonlinear materials using gold nanoantenna array}},\ }\href
  {https://doi.org/10.1038/s41598-017-19066-3} {\bibfield  {journal} {\bibinfo
  {journal} {Scientific Reports}\ }\textbf {\bibinfo {volume} {8}},\ \bibinfo
  {pages} {780} (\bibinfo {year} {2018})}\BibitemShut {NoStop}%
\bibitem [{\citenamefont {Postaci}\ \emph {et~al.}(2018)\citenamefont
  {Postaci}, \citenamefont {Yildiz}, \citenamefont {Bek},\ and\ \citenamefont
  {Tasgin}}]{Postaci2018SilentIntensities}%
  \BibitemOpen
  \bibfield  {author} {\bibinfo {author} {\bibfnamefont {S.}~\bibnamefont
  {Postaci}}, \bibinfo {author} {\bibfnamefont {B.~C.}\ \bibnamefont {Yildiz}},
  \bibinfo {author} {\bibfnamefont {A.}~\bibnamefont {Bek}}, and\ \bibinfo
  {author} {\bibfnamefont {M.~E.}\ \bibnamefont {Tasgin}},\ }\bibfield  {title}
  {\bibinfo {title} {{Silent enhancement of SERS signal without increasing hot
  spot intensities}},\ }\href {https://doi.org/10.1515/nanoph-2018-0089}
  {\bibfield  {journal} {\bibinfo  {journal} {Nanophotonics}\ }\textbf
  {\bibinfo {volume} {7}},\ \bibinfo {pages} {1687} (\bibinfo {year}
  {2018})}\BibitemShut {NoStop}%
\bibitem [{\citenamefont {Melikyan}\ and\ \citenamefont
  {Minassian}(2004)}]{Melikyan2004OnNanoparticles}%
  \BibitemOpen
  \bibfield  {author} {\bibinfo {author} {\bibfnamefont {A.}~\bibnamefont
  {Melikyan}}and\ \bibinfo {author} {\bibfnamefont {H.}~\bibnamefont
  {Minassian}},\ }\bibfield  {title} {\bibinfo {title} {{On surface plasmon
  damping in metallic nanoparticles}},\ }\href
  {https://doi.org/10.1007/s00340-004-1403-z} {\bibfield  {journal} {\bibinfo
  {journal} {Applied Physics B}\ }\textbf {\bibinfo {volume} {78}},\ \bibinfo
  {pages} {453} (\bibinfo {year} {2004})}\BibitemShut {NoStop}%
\bibitem [{\citenamefont {Kirakosyan}\ \emph {et~al.}(2016)\citenamefont
  {Kirakosyan}, \citenamefont {Stockman},\ and\ \citenamefont
  {Shahbazyan}}]{Kirakosyan2016SurfaceNanoshells}%
  \BibitemOpen
  \bibfield  {author} {\bibinfo {author} {\bibfnamefont {A.~S.}\ \bibnamefont
  {Kirakosyan}}, \bibinfo {author} {\bibfnamefont {M.~I.}\ \bibnamefont
  {Stockman}}, and\ \bibinfo {author} {\bibfnamefont {T.~V.}\ \bibnamefont
  {Shahbazyan}},\ }\bibfield  {title} {\bibinfo {title} {{Surface plasmon
  lifetime in metal nanoshells}},\ }\href
  {https://doi.org/10.1103/PhysRevB.94.155429} {\bibfield  {journal} {\bibinfo
  {journal} {Physical Review B}\ }\textbf {\bibinfo {volume} {94}},\ \bibinfo
  {pages} {155429} (\bibinfo {year} {2016})}\BibitemShut {NoStop}%
\bibitem [{\citenamefont {Mahan}(2018)}]{Mahan2018LifetimePlasmons}%
  \BibitemOpen
  \bibfield  {author} {\bibinfo {author} {\bibfnamefont {G.~D.}\ \bibnamefont
  {Mahan}},\ }\bibfield  {title} {\bibinfo {title} {{Lifetime of surface
  plasmons}},\ }\href {https://doi.org/10.1103/PhysRevB.97.075405} {\bibfield
  {journal} {\bibinfo  {journal} {Physical Review B}\ }\textbf {\bibinfo
  {volume} {97}},\ \bibinfo {pages} {075405} (\bibinfo {year}
  {2018})}\BibitemShut {NoStop}%
\bibitem [{\citenamefont {Di~Vece}(2018)}]{DiVece2018VeryParticles}%
  \BibitemOpen
  \bibfield  {author} {\bibinfo {author} {\bibfnamefont {M.}~\bibnamefont
  {Di~Vece}},\ }\bibfield  {title} {\bibinfo {title} {{Very Long Plasmon
  Oscillation Lifetimes in the Gap Between Two Gold Particles}},\ }\href
  {https://doi.org/10.1007/s11468-017-0640-z} {\bibfield  {journal} {\bibinfo
  {journal} {Plasmonics}\ }\textbf {\bibinfo {volume} {13}},\ \bibinfo {pages}
  {1367} (\bibinfo {year} {2018})}\BibitemShut {NoStop}%
\bibitem [{\citenamefont {Chapkin}\ \emph {et~al.}(2018)\citenamefont
  {Chapkin}, \citenamefont {Bursi}, \citenamefont {Stec}, \citenamefont
  {Lauchner}, \citenamefont {Hogan}, \citenamefont {Cui}, \citenamefont
  {Nordlander},\ and\ \citenamefont {Halas}}]{Chapkin2018LifetimeLimit.}%
  \BibitemOpen
  \bibfield  {author} {\bibinfo {author} {\bibfnamefont {K.~D.}\ \bibnamefont
  {Chapkin}}, \bibinfo {author} {\bibfnamefont {L.}~\bibnamefont {Bursi}},
  \bibinfo {author} {\bibfnamefont {G.~J.}\ \bibnamefont {Stec}}, \bibinfo
  {author} {\bibfnamefont {A.}~\bibnamefont {Lauchner}}, \bibinfo {author}
  {\bibfnamefont {N.~J.}\ \bibnamefont {Hogan}}, \bibinfo {author}
  {\bibfnamefont {Y.}~\bibnamefont {Cui}}, \bibinfo {author} {\bibfnamefont
  {P.}~\bibnamefont {Nordlander}}, and\ \bibinfo {author} {\bibfnamefont
  {N.~J.}\ \bibnamefont {Halas}},\ }\bibfield  {title} {\bibinfo {title}
  {{Lifetime dynamics of plasmons in the few-atom limit.}},\ }\href
  {https://doi.org/10.1073/pnas.1805357115} {\bibfield  {journal} {\bibinfo
  {journal} {Proceedings of the National Academy of Sciences of the United
  States of America}\ }\textbf {\bibinfo {volume} {115}},\ \bibinfo {pages}
  {9134} (\bibinfo {year} {2018})}\BibitemShut {NoStop}%
\bibitem [{\citenamefont {Noginov}\ \emph {et~al.}(2009)\citenamefont
  {Noginov}, \citenamefont {Zhu}, \citenamefont {Belgrave}, \citenamefont
  {Bakker}, \citenamefont {Shalaev}, \citenamefont {Narimanov}, \citenamefont
  {Stout}, \citenamefont {Herz}, \citenamefont {Suteewong},\ and\ \citenamefont
  {Wiesner}}]{Noginov2009DemonstrationNanolaser}%
  \BibitemOpen
  \bibfield  {author} {\bibinfo {author} {\bibfnamefont {M.~A.}\ \bibnamefont
  {Noginov}}, \bibinfo {author} {\bibfnamefont {G.}~\bibnamefont {Zhu}},
  \bibinfo {author} {\bibfnamefont {A.~M.}\ \bibnamefont {Belgrave}}, \bibinfo
  {author} {\bibfnamefont {R.}~\bibnamefont {Bakker}}, \bibinfo {author}
  {\bibfnamefont {V.~M.}\ \bibnamefont {Shalaev}}, \bibinfo {author}
  {\bibfnamefont {E.~E.}\ \bibnamefont {Narimanov}}, \bibinfo {author}
  {\bibfnamefont {S.}~\bibnamefont {Stout}}, \bibinfo {author} {\bibfnamefont
  {E.}~\bibnamefont {Herz}}, \bibinfo {author} {\bibfnamefont {T.}~\bibnamefont
  {Suteewong}}, and\ \bibinfo {author} {\bibfnamefont {U.}~\bibnamefont
  {Wiesner}},\ }\bibfield  {title} {\bibinfo {title} {{Demonstration of a
  spaser-based nanolaser}},\ }\href {https://doi.org/10.1038/nature08318}
  {\bibfield  {journal} {\bibinfo  {journal} {Nature}\ }\textbf {\bibinfo
  {volume} {460}},\ \bibinfo {pages} {1110} (\bibinfo {year}
  {2009})}\BibitemShut {NoStop}%
\bibitem [{\citenamefont {Tasgin}(2013)}]{Tasgin2013MetalLifetime}%
  \BibitemOpen
  \bibfield  {author} {\bibinfo {author} {\bibfnamefont {M.~E.}\ \bibnamefont
  {Tasgin}},\ }\bibfield  {title} {\bibinfo {title} {{Metal nanoparticle
  plasmons operating within a quantum lifetime}},\ }\href
  {https://doi.org/10.1039/c3nr02270f} {\bibfield  {journal} {\bibinfo
  {journal} {Nanoscale}\ }\textbf {\bibinfo {volume} {5}},\ \bibinfo {pages}
  {8616} (\bibinfo {year} {2013})}\BibitemShut {NoStop}%
\bibitem [{\citenamefont {Stockman}(2010)}]{Stockman2010Dark-hotResonances}%
  \BibitemOpen
  \bibfield  {author} {\bibinfo {author} {\bibfnamefont {M.~I.}\ \bibnamefont
  {Stockman}},\ }\bibfield  {title} {\bibinfo {title} {{Dark-hot resonances}},\
  }\href {https://doi.org/10.1038/467541a} {\bibfield  {journal} {\bibinfo
  {journal} {Nature}\ }\textbf {\bibinfo {volume} {467}},\ \bibinfo {pages}
  {541} (\bibinfo {year} {2010})}\BibitemShut {NoStop}%
\bibitem [{\citenamefont {G{\'{o}}mez}\ \emph {et~al.}(2013)\citenamefont
  {G{\'{o}}mez}, \citenamefont {Teo}, \citenamefont {Altissimo}, \citenamefont
  {Davis}, \citenamefont {Earl},\ and\ \citenamefont
  {Roberts}}]{Gomez2013ThePlasmonics}%
  \BibitemOpen
  \bibfield  {author} {\bibinfo {author} {\bibfnamefont {D.~E.}\ \bibnamefont
  {G{\'{o}}mez}}, \bibinfo {author} {\bibfnamefont {Z.~Q.}\ \bibnamefont
  {Teo}}, \bibinfo {author} {\bibfnamefont {M.}~\bibnamefont {Altissimo}},
  \bibinfo {author} {\bibfnamefont {T.~J.}\ \bibnamefont {Davis}}, \bibinfo
  {author} {\bibfnamefont {S.}~\bibnamefont {Earl}}, and\ \bibinfo {author}
  {\bibfnamefont {A.}~\bibnamefont {Roberts}},\ }\bibfield  {title} {\bibinfo
  {title} {{The Dark Side of Plasmonics}},\ }\href
  {https://doi.org/10.1021/nl401656e} {\bibfield  {journal} {\bibinfo
  {journal} {Nano Letters}\ }\textbf {\bibinfo {volume} {13}},\ \bibinfo
  {pages} {3722} (\bibinfo {year} {2013})}\BibitemShut {NoStop}%
\bibitem [{\citenamefont {Panaro}\ \emph {et~al.}(2014)\citenamefont {Panaro},
  \citenamefont {Nazir}, \citenamefont {Liberale}, \citenamefont {Das},
  \citenamefont {Wang}, \citenamefont {De~Angelis}, \citenamefont
  {Proietti~Zaccaria}, \citenamefont {Di~Fabrizio},\ and\ \citenamefont
  {Toma}}]{Panaro2014DarkApproach}%
  \BibitemOpen
  \bibfield  {author} {\bibinfo {author} {\bibfnamefont {S.}~\bibnamefont
  {Panaro}}, \bibinfo {author} {\bibfnamefont {A.}~\bibnamefont {Nazir}},
  \bibinfo {author} {\bibfnamefont {C.}~\bibnamefont {Liberale}}, \bibinfo
  {author} {\bibfnamefont {G.}~\bibnamefont {Das}}, \bibinfo {author}
  {\bibfnamefont {H.}~\bibnamefont {Wang}}, \bibinfo {author} {\bibfnamefont
  {F.}~\bibnamefont {De~Angelis}}, \bibinfo {author} {\bibfnamefont
  {R.}~\bibnamefont {Proietti~Zaccaria}}, \bibinfo {author} {\bibfnamefont
  {E.}~\bibnamefont {Di~Fabrizio}}, and\ \bibinfo {author} {\bibfnamefont
  {A.}~\bibnamefont {Toma}},\ }\bibfield  {title} {\bibinfo {title} {{Dark to
  Bright Mode Conversion on Dipolar Nanoantennas: A Symmetry-Breaking
  Approach}},\ }\href {https://doi.org/10.1021/ph500044w} {\bibfield  {journal}
  {\bibinfo  {journal} {ACS Photonics}\ }\textbf {\bibinfo {volume} {1}},\
  \bibinfo {pages} {310} (\bibinfo {year} {2014})}\BibitemShut {NoStop}%
\bibitem [{\citenamefont {Gao}\ \emph {et~al.}(2018)\citenamefont {Gao},
  \citenamefont {Zhou}, \citenamefont {Shi}, \citenamefont {Guo},\ and\
  \citenamefont {Tong}}]{Gao:18}%
  \BibitemOpen
  \bibfield  {author} {\bibinfo {author} {\bibfnamefont {Y.}~\bibnamefont
  {Gao}}, \bibinfo {author} {\bibfnamefont {N.}~\bibnamefont {Zhou}}, \bibinfo
  {author} {\bibfnamefont {Z.}~\bibnamefont {Shi}}, \bibinfo {author}
  {\bibfnamefont {X.}~\bibnamefont {Guo}}, and\ \bibinfo {author}
  {\bibfnamefont {L.}~\bibnamefont {Tong}},\ }\bibfield  {title} {\bibinfo
  {title} {Dark dimer mode excitation and strong coupling with a nanorod
  dipole},\ }\href {https://doi.org/10.1364/PRJ.6.000887} {\bibfield  {journal}
  {\bibinfo  {journal} {Photon. Res.}\ }\textbf {\bibinfo {volume} {6}},\
  \bibinfo {pages} {887} (\bibinfo {year} {2018})}\BibitemShut {NoStop}%
\bibitem [{\citenamefont {Taflove}\ and\ \citenamefont
  {Hagness}(2005)}]{Taflove2005ComputationalMethod}%
  \BibitemOpen
  \bibfield  {author} {\bibinfo {author} {\bibfnamefont {A.}~\bibnamefont
  {Taflove}}and\ \bibinfo {author} {\bibfnamefont {S.~C.}\ \bibnamefont
  {Hagness}},\ }\href@noop {} {\emph {\bibinfo {title} {{Computational
  electrodynamics: the finite-difference time-domain method}}}}\ (\bibinfo
  {publisher} {Artech House},\ \bibinfo {year} {2005})\ p.\ \bibinfo {pages}
  {1006}\BibitemShut {NoStop}%
\bibitem [{\citenamefont {Zhang}\ and\ \citenamefont {Xu}(2016)}]{C6NR03806A}%
  \BibitemOpen
  \bibfield  {author} {\bibinfo {author} {\bibfnamefont {S.}~\bibnamefont
  {Zhang}}and\ \bibinfo {author} {\bibfnamefont {H.}~\bibnamefont {Xu}},\
  }\bibfield  {title} {\bibinfo {title} {Tunable dark plasmons in a metallic
  nanocube dimer: toward ultimate sensitivity nanoplasmonic sensors},\ }\href
  {https://doi.org/10.1039/C6NR03806A} {\bibfield  {journal} {\bibinfo
  {journal} {Nanoscale}\ }\textbf {\bibinfo {volume} {8}},\ \bibinfo {pages}
  {13722} (\bibinfo {year} {2016})}\BibitemShut {NoStop}%
\bibitem [{\citenamefont {Tasgin}\ \emph {et~al.}(2018)\citenamefont {Tasgin},
  \citenamefont {Bek},\ and\ \citenamefont {Postaci}}]{Tasgin2018FanoResponse}%
  \BibitemOpen
  \bibfield  {author} {\bibinfo {author} {\bibfnamefont {M.~E.}\ \bibnamefont
  {Tasgin}}, \bibinfo {author} {\bibfnamefont {A.}~\bibnamefont {Bek}}, and\
  \bibinfo {author} {\bibfnamefont {S.}~\bibnamefont {Postaci}},\ }\bibfield
  {title} {\bibinfo {title} {{Fano Resonances in the Linear and Nonlinear
  Plasmonic Response}}\ }(\bibinfo  {publisher} {Springer, Cham},\ \bibinfo
  {year} {2018})\ pp.\ \bibinfo {pages} {1--31}\BibitemShut {NoStop}%
\bibitem [{\citenamefont {{P. Nordlander}}\ \emph {et~al.}(2004)\citenamefont
  {{P. Nordlander}}, \citenamefont {Oubre}, \citenamefont {Prodan},
  \citenamefont {{K. Li}},\ and\ \citenamefont
  {Stockman}}]{Nordlander2004PlasmonDimers}%
  \BibitemOpen
  \bibfield  {author} {\bibinfo {author} {\bibnamefont {{P. Nordlander}}},
  \bibinfo {author} {\bibfnamefont {C.}~\bibnamefont {Oubre}}, \bibinfo
  {author} {\bibfnamefont {E.}~\bibnamefont {Prodan}}, \bibinfo {author}
  {\bibnamefont {{K. Li}}}, and\ \bibinfo {author} {\bibfnamefont {M.~I.}\
  \bibnamefont {Stockman}},\ }\bibfield  {title} {\bibinfo {title} {{Plasmon
  Hybridization in Nanoparticle Dimers}},\ }\bibfield  {journal} {\bibinfo
  {journal} {Nano Letters}\ }\href {https://doi.org/10.1021/NL049681C}
  {10.1021/NL049681C} (\bibinfo {year} {2004})\BibitemShut {NoStop}%
\bibitem [{\citenamefont {Sun}\ \emph {et~al.}(2008)\citenamefont {Sun},
  \citenamefont {Al-Amri}, \citenamefont {Kamli},\ and\ \citenamefont
  {Zubairy}}]{Sun2008LambModes}%
  \BibitemOpen
  \bibfield  {author} {\bibinfo {author} {\bibfnamefont {Q.}~\bibnamefont
  {Sun}}, \bibinfo {author} {\bibfnamefont {M.}~\bibnamefont {Al-Amri}},
  \bibinfo {author} {\bibfnamefont {A.}~\bibnamefont {Kamli}}, and\ \bibinfo
  {author} {\bibfnamefont {M.~S.}\ \bibnamefont {Zubairy}},\ }\bibfield
  {title} {\bibinfo {title} {{Lamb shift due to surface plasmon polariton
  modes}},\ }\href {https://doi.org/10.1103/PhysRevA.77.062501} {\bibfield
  {journal} {\bibinfo  {journal} {Physical Review A}\ }\textbf {\bibinfo
  {volume} {77}},\ \bibinfo {pages} {062501} (\bibinfo {year}
  {2008})}\BibitemShut {NoStop}%
\bibitem [{\citenamefont {Yannopapas}\ \emph {et~al.}(2009)\citenamefont
  {Yannopapas}, \citenamefont {Paspalakis},\ and\ \citenamefont
  {Vitanov}}]{Yannopapas2009Plasmon-InducedNanostructures}%
  \BibitemOpen
  \bibfield  {author} {\bibinfo {author} {\bibfnamefont {V.}~\bibnamefont
  {Yannopapas}}, \bibinfo {author} {\bibfnamefont {E.}~\bibnamefont
  {Paspalakis}}, and\ \bibinfo {author} {\bibfnamefont {N.~V.}\ \bibnamefont
  {Vitanov}},\ }\bibfield  {title} {\bibinfo {title} {{Plasmon-Induced
  Enhancement of Quantum Interference near Metallic Nanostructures}},\ }\href
  {https://doi.org/10.1103/PhysRevLett.103.063602} {\bibfield  {journal}
  {\bibinfo  {journal} {Physical Review Letters}\ }\textbf {\bibinfo {volume}
  {103}},\ \bibinfo {pages} {063602} (\bibinfo {year} {2009})}\BibitemShut
  {NoStop}%
\bibitem [{\citenamefont {Qurban}\ \emph {et~al.}(2018)\citenamefont {Qurban},
  \citenamefont {Ikram}, \citenamefont {Ge},\ and\ \citenamefont
  {Zubairy}}]{Qurban2018EntanglementNanoring}%
  \BibitemOpen
  \bibfield  {author} {\bibinfo {author} {\bibfnamefont {M.}~\bibnamefont
  {Qurban}}, \bibinfo {author} {\bibfnamefont {M.}~\bibnamefont {Ikram}},
  \bibinfo {author} {\bibfnamefont {G.-Q.}\ \bibnamefont {Ge}}, and\ \bibinfo
  {author} {\bibfnamefont {M.~S.}\ \bibnamefont {Zubairy}},\ }\bibfield
  {title} {\bibinfo {title} {{Entanglement generation among quantum dots and
  surface plasmons of a metallic nanoring}},\ }\href
  {https://doi.org/10.1088/1361-6455/aacbfc} {\bibfield  {journal} {\bibinfo
  {journal} {Journal of Physics B: Atomic, Molecular and Optical Physics}\
  }\textbf {\bibinfo {volume} {51}},\ \bibinfo {pages} {155502} (\bibinfo
  {year} {2018})}\BibitemShut {NoStop}%
\bibitem [{\citenamefont {Trugler}(2016)}]{Andreas2016}%
  \BibitemOpen
  \bibfield  {author} {\bibinfo {author} {\bibfnamefont {A.}~\bibnamefont
  {Trugler}},\ }\href@noop {} {\emph {\bibinfo {title} {Optical Properties of
  Metallic Nanoparticles}}}\ (\bibinfo  {publisher} {Springer},\ \bibinfo
  {year} {2016})\BibitemShut {NoStop}%
\bibitem [{\citenamefont {Zoric}\ \emph {et~al.}(2011)\citenamefont {Zoric},
  \citenamefont {Zach}, \citenamefont {Kasemo},\ and\ \citenamefont
  {Langhammer}}]{doi:10.1021/nn102166t}%
  \BibitemOpen
  \bibfield  {author} {\bibinfo {author} {\bibfnamefont {I.}~\bibnamefont
  {Zoric}}, \bibinfo {author} {\bibfnamefont {M.}~\bibnamefont {Zach}},
  \bibinfo {author} {\bibfnamefont {B.}~\bibnamefont {Kasemo}}, and\ \bibinfo
  {author} {\bibfnamefont {C.}~\bibnamefont {Langhammer}},\ }\bibfield  {title}
  {\bibinfo {title} {Gold, platinum, and aluminum nanodisk plasmons: Material
  independence, subradiance, and damping mechanisms},\ }\href
  {https://doi.org/10.1021/nn102166t} {\bibfield  {journal} {\bibinfo
  {journal} {ACS Nano}\ }\textbf {\bibinfo {volume} {5}},\ \bibinfo {pages}
  {2535} (\bibinfo {year} {2011})},\ \bibinfo {note} {pMID:
  21438568}\BibitemShut {NoStop}%
\bibitem [{\citenamefont {Verellen}\ \emph {et~al.}(2011)\citenamefont
  {Verellen}, \citenamefont {Dorpe}, \citenamefont {Vercruysse}, \citenamefont
  {Vandenbosch},\ and\ \citenamefont {Moshchalkov}}]{Verellen:11}%
  \BibitemOpen
  \bibfield  {author} {\bibinfo {author} {\bibfnamefont {N.}~\bibnamefont
  {Verellen}}, \bibinfo {author} {\bibfnamefont {P.~V.}\ \bibnamefont {Dorpe}},
  \bibinfo {author} {\bibfnamefont {D.}~\bibnamefont {Vercruysse}}, \bibinfo
  {author} {\bibfnamefont {G.~A.~E.}\ \bibnamefont {Vandenbosch}}, and\
  \bibinfo {author} {\bibfnamefont {V.~V.}\ \bibnamefont {Moshchalkov}},\
  }\bibfield  {title} {\bibinfo {title} {Dark and bright localized surface
  plasmons in nanocrosses},\ }\href {https://doi.org/10.1364/OE.19.011034}
  {\bibfield  {journal} {\bibinfo  {journal} {Opt. Express}\ }\textbf {\bibinfo
  {volume} {19}},\ \bibinfo {pages} {11034} (\bibinfo {year}
  {2011})}\BibitemShut {NoStop}%
\bibitem [{\citenamefont {Johnson}\ and\ \citenamefont
  {Christy}(1972)}]{Johnson1972}%
  \BibitemOpen
  \bibfield  {author} {\bibinfo {author} {\bibfnamefont {P.~B.}\ \bibnamefont
  {Johnson}}and\ \bibinfo {author} {\bibfnamefont {R.~W.}\ \bibnamefont
  {Christy}},\ }\bibfield  {title} {\bibinfo {title} {{Optical Constants of the
  Noble Metals}},\ }\href {https://doi.org/10.1103/PhysRevB.6.4370} {\bibfield
  {journal} {\bibinfo  {journal} {Physical Review B}\ }\textbf {\bibinfo
  {volume} {6}},\ \bibinfo {pages} {4370} (\bibinfo {year} {1972})}\BibitemShut
  {NoStop}%
\bibitem [{\citenamefont {Chen}\ \emph {et~al.}(2013)\citenamefont {Chen},
  \citenamefont {Shao}, \citenamefont {Li},\ and\ \citenamefont
  {Wang}}]{C2CS35367A}%
  \BibitemOpen
  \bibfield  {author} {\bibinfo {author} {\bibfnamefont {H.}~\bibnamefont
  {Chen}}, \bibinfo {author} {\bibfnamefont {L.}~\bibnamefont {Shao}}, \bibinfo
  {author} {\bibfnamefont {Q.}~\bibnamefont {Li}}, and\ \bibinfo {author}
  {\bibfnamefont {J.}~\bibnamefont {Wang}},\ }\bibfield  {title} {\bibinfo
  {title} {Gold nanorods and their plasmonic properties},\ }\href
  {https://doi.org/10.1039/C2CS35367A} {\bibfield  {journal} {\bibinfo
  {journal} {Chem. Soc. Rev.}\ }\textbf {\bibinfo {volume} {42}},\ \bibinfo
  {pages} {2679} (\bibinfo {year} {2013})}\BibitemShut {NoStop}%
\bibitem [{\citenamefont {Langhammer}\ \emph {et~al.}(2006)\citenamefont
  {Langhammer}, \citenamefont {Yuan}, \citenamefont {ZoriÄ‡},\ and\
  \citenamefont {Kasemo}}]{doi:10.1021/nl060219x}%
  \BibitemOpen
  \bibfield  {author} {\bibinfo {author} {\bibfnamefont {C.}~\bibnamefont
  {Langhammer}}, \bibinfo {author} {\bibfnamefont {Z.}~\bibnamefont {Yuan}},
  \bibinfo {author} {\bibfnamefont {I.}~\bibnamefont {ZoriÄ‡}}, and\
  \bibinfo {author} {\bibfnamefont {B.}~\bibnamefont {Kasemo}},\ }\bibfield
  {title} {\bibinfo {title} {Plasmonic properties of supported pt and pd
  nanostructures},\ }\href {https://doi.org/10.1021/nl060219x} {\bibfield
  {journal} {\bibinfo  {journal} {Nano Letters}\ }\textbf {\bibinfo {volume}
  {6}},\ \bibinfo {pages} {833} (\bibinfo {year} {2006})},\ \bibinfo {note}
  {pMID: 16608293}\BibitemShut {NoStop}%
\bibitem [{\citenamefont {Palik}(1998)}]{Palik1998HandbookSolids}%
  \BibitemOpen
  \bibfield  {author} {\bibinfo {author} {\bibfnamefont {E.~D.}\ \bibnamefont
  {Palik}},\ }\href
  {https://www.sciencedirect.com/book/9780125444156/handbook-of-optical-constants-of-solids}
  {\emph {\bibinfo {title} {{Handbook of optical constants of solids}}}}\
  (\bibinfo  {publisher} {Academic Press},\ \bibinfo {year} {1998})\BibitemShut
  {NoStop}%
\bibitem [{\citenamefont {and}\ and\ \citenamefont
  {El-Sayed*}(1999)}]{and1999SpectralNanorods}%
  \BibitemOpen
  \bibfield  {author} {\bibinfo {author} {\bibfnamefont {S.~L.}\ \bibnamefont
  {and}}and\ \bibinfo {author} {\bibfnamefont {M.~A.}\ \bibnamefont
  {El-Sayed*}},\ }\bibfield  {title} {\bibinfo {title} {{Spectral Properties
  and Relaxation Dynamics of Surface Plasmon Electronic Oscillations in Gold
  and Silver Nanodots and Nanorods}},\ }\bibfield  {journal} {\bibinfo
  {journal} {The Journal of Physical Chemistry B}\ }\href
  {https://doi.org/10.1021/JP9917648} {10.1021/JP9917648} (\bibinfo {year}
  {1999})\BibitemShut {NoStop}%
\bibitem [{\citenamefont {Anger}\ \emph {et~al.}(2006)\citenamefont {Anger},
  \citenamefont {Bharadwaj},\ and\ \citenamefont
  {Novotny}}]{Anger2006EnhancementFluorescence}%
  \BibitemOpen
  \bibfield  {author} {\bibinfo {author} {\bibfnamefont {P.}~\bibnamefont
  {Anger}}, \bibinfo {author} {\bibfnamefont {P.}~\bibnamefont {Bharadwaj}},
  and\ \bibinfo {author} {\bibfnamefont {L.}~\bibnamefont {Novotny}},\
  }\bibfield  {title} {\bibinfo {title} {{Enhancement and Quenching of
  Single-Molecule Fluorescence}},\ }\href
  {https://doi.org/10.1103/PhysRevLett.96.113002} {\bibfield  {journal}
  {\bibinfo  {journal} {Physical Review Letters}\ }\textbf {\bibinfo {volume}
  {96}},\ \bibinfo {pages} {113002} (\bibinfo {year} {2006})}\BibitemShut
  {NoStop}%
\end{thebibliography}%

\end{document}